\documentclass[aps,preprint,11pt,preprintnumbers,nofootinbib,superscriptaddress]{revtex4}

\usepackage{graphicx}
\usepackage{dcolumn}
\usepackage{bm}
\usepackage{amssymb}
\usepackage{amsmath}
\usepackage{graphicx}
\usepackage{xcolor}
\usepackage{slashed}
\newcommand*{\pd}{\partial}
\newcommand*{\ra}{\rightarrow}
\newcommand*{\eps}{\epsilon}
\newcommand*{\half}{\frac{1}{2}}

\begin{document}

\title{Vortex in holographic two-band superfluid/superconductor}
\author{Mu-Sheng Wu}
\email{msgn123@gmail.com}
\affiliation{Department of Physics and Astronomy, University of Alabama, Tuscaloosa, AL 35487, USA}
\affiliation{National Center of Theoretical Sciences, National Tsing Hua University, Hsinchu, Taiwan 300}
\author{Shang-Yu Wu}
\email{loganwu@gmail.com}
\affiliation{Department of Electrophysics and Shing-Tung Yau Center, National Chiao Tung University, Hsinchu, Taiwan 300}
\author{Hai-Qing Zhang}
\email{H.Q.Zhang@uu.nl}
\affiliation{Institute for Theoretical Physics, Utrecht University, Leuvenlaan 4, 3584 CE Utrecht, The Netherlands}


\begin{abstract}
We construct numerically static vortex solutions in a holographic model of two-band superconductor with an interband Josephson coupling in both the superfluid and superconductor regime. We investigate the effects of the interband coupling on the order parameter of each superconducting band in the vortex solution, and we find that it is different for each of the two bands. We compute also the free energy, critical magnetic field, magnetic penetration length and coherence lengths for the two bands, and we study their dependence on the interband coupling and temperature. Interestingly, we find that the coherence lengths of the two bands are close to identical.
\end{abstract}

\maketitle

\section{Introduction}
Multiband superconductors have attracted much attention since the discovery of the first two-band superconductor in MgB$_{2}$~\cite{MgB2}, and more recently the first iron-based superconductor~\cite{iron}. Many novel features were discovered in MgB$_{2}$, such as having a large critical current, an anisotropy in the Fermi velocity, and an anomalous vortex dynamics~\cite{MgB2 review}, while iron-based superconductors may give rise to a new class of high-temperature superconductors, given the similarity of their planar structures and their phase diagrams to the cuprates~\cite{FeSC review}.

Multiband superconductivity has been studied using Ginzburg-Landau (GL) theory now generalized from having just a single superconductor condensate to having multiple ones \cite{TCGL}. Interesting phenomena such as the formation of interband phase difference soliton~\cite{fractional flux quanta}, fractional flux quanta~\cite{fractional flux quanta}, and possibly type-1.5 superconductivity~\cite{type 1.5} in which vortex clusters can coexist with the Meissner domain are seen. Our goal in this paper is to study multiband superconductivity beyond the regime of validity of the GL theory, i.e. fields are not assumed to be small. In particular, we shall study vortices in two-band superconductors that are strongly coupled.

The tool we use to study strongly-coupled/correlated systems is the AdS/CFT correspondence~\cite{Maldacena:1997re,Witten:1998qj} or ``holography'', which has proven to be very useful in a variety of different areas, including QCD~\cite{QCD}, heavy ion physics~\cite{heavy ion collision}, and superconductivity~\cite{Hartnoll:2008vx, Hartnoll:2008kx, Gubser:2008wv, Chen:2010mk, Benini:2010pr, Kim:2013oba}. In Ref.~\cite{Wen:2013ufa}, a holographic model of two-band superconductor was constructed~\footnote{Other examples of holographic model of multiband superconductor are given in Refs.~\cite{Huang:2011ac,Krikun:2012yj}.}. The model takes into account fully the back-reaction from the matter sector on the gravity background, and emphasizes the effects of the interband Josephson coupling, which was realized by a Josephson-like coupling between two bulk complex scalar fields. The transport properties of the holographic model were studied, and were shown to have the same qualitative features
as seen in experiments.

In this paper, we continue the study of two-band superconductor initiated in Ref.~\cite{Wen:2013ufa} by two of the present authors. In particular, we shall look for vortex solutions as a response to the magnetic field. It is known that the type of AdS-boundary conditions imposed on the bulk $U(1)$ gauge field determines the kind of vortices found: Dirichlet type give rises to superfluid vortices, Neumann type to superconductor vortices~\cite{Domenech:2010nf,Dias:2013bwa}. Here we shall study both types. We shall also check whether the purported type-1.5 superconductivity -- which were seen in some studies, but not all -- exists in our holographic model. A quantitative indicator for a type-1.5 two-band superconductor is when the coherence lengths for the two bands, $\xi_1$ and $\xi_2$, and the magnetic penetration length, $\lambda$, satisfy the relation $\xi_1 < \sqrt{2}\lambda < \xi_2$~\cite{type 1.5}. By extracting the coherence and penetration lengths from our holographic model, we can test for type-1.5
superconductivity.

The paper is organized as follows. In Sec.~\ref{sec:model}, we describe the set-up for finding vortex solutions in the holographic model of two-band superconductor of Ref.~\cite{Wen:2013ufa}. We give the ansatz for the vortex solution and we specify the boundary conditions for both superfluid and superconductor type vortices. In Sec.~\ref{sec:vortices}, we  study the vortex solutions in detail, and we compute the coherence lengths for both types of vortices. In the case of superconductor vortex, we compute also the magnetic penetration length as the magnetic field is dynamical, showing that the holographic two-band superconductor is always type II. We conclude in Sec.~\ref{sec:concl} with a summary.

\section{\label{sec:model} The Holographic Model}
We consider the minimal holographic model of two-band superconductor in $AdS_4$ given in Ref.~\cite{Wen:2013ufa}:
\begin{gather}\label{eq:action}
S = \frac{1}{2\kappa^2}\int\!d^4x\sqrt{-g}\left[
R + \frac{6}{L^{2}}-\frac{1}{4}F^{2}-|\partial\psi_{1}-iqA\psi_{1}|^{2}-|\partial\psi_{2}-iqA\psi_{2}|^{2}-V(\psi_{1},\psi_{2})\right] \,, \\
V(\psi_{1},\psi_{2}) =
m_{1}^{2}|\psi_{1}|^{2}+m_{2}^{2}|\psi_{2}|^{2}+\epsilon(\psi_{1}\psi^{*}_{2}+\psi^{*}_{1}\psi_{2})+\eta|\psi_{1}|^{2}|\psi_{2}|^{2} \,,
\end{gather}
where $\psi_{1,2}$ are complex scalar fields with masses $m_{1,2}$ respectively, $A_\mu$ is the $U(1)$ gauge field with
$F = dA$ the field strength, and $q$ is the $U(1)$ charge of the complex scalar fields~\footnote{Given the form of the  interactions (the quadratic ones in particular) in our action, gauge invariance requires the scalars to have the same charge.}. In the potential $V$, $\epsilon$ denotes an interband Josephson coupling, and $\eta$ a density-density coupling.

To look for vortex solutions, it is more convenient to write the complex scalars as $\psi_{1}=\varphi_{1}e^{i\theta_{1}}$ and $\psi_{2}=\varphi_{2}e^{i\theta_{2}}$ in terms of their moduli $\varphi_{1,2}$ and phases $\theta_{1,2}$. The action then becomes
\begin{align}\label{eq:action2}
S &=\frac{1}{2\kappa^{2}}\int\!d^{4}x\sqrt{-g}\bigg[R+\frac{6}{L^{2}}-\frac{1}{4}F^{2} \notag \\
&\qquad\qquad -(\partial\varphi_{1})^{2}-\varphi_{1}^{2}(\partial_{\mu}\theta_{1}-qA_{\mu})^{2}
-(\partial\varphi_{2})^{2}-\varphi_{2}^{2}(\partial_{\mu}\theta_{2}-qA_{\mu})^{2}-V(\varphi_{1},\varphi_{2})\bigg] \,,
\end{align}
with
\begin{equation}
V(\varphi_{1},\varphi_{2}) = m^2_{1}\varphi_{1}^{2} + m^2_{2}\varphi_{2}^{2} + 2\epsilon\varphi_{1}\varphi_{2}\cos(\theta_{1}-\theta_{2}) + \eta\varphi_{1}^{2}\varphi_{2}^{2} \,.
\end{equation}
We shall work in the probe limit, where the matter sector does not cause backreaction on the background metric. We take the background to be an AdS-Schwarzchild black hole, whose metric is given by
\begin{equation}
ds^{2}=\frac{1}{z^{2}}\left(-f(z)dt^{2}+\frac{dz^{2}}{f(z)}+d\rho^{2}+\rho^{2}d\phi^{2}\right), \quad f(r)=1-\left(\frac{z}{z_h}\right)^{3},
\end{equation}
where $z_{h}$ is the location of horizon. For convenience, we have used polar coordinates $(\rho,\phi)$ for the two-dimensional (2D) plane in the spatial field theory directions.

\subsection{The vortex solution}
A consistent ansatz respecting the global $U(1)$ symmetry and rotational symmetry on the 2D plane is given by
\begin{gather}
\varphi_i=\varphi_{i}(\rho,z) \,, \quad \theta_i = n_i\phi \,, \quad i = 1,\,2 \,, \\
A_t = A_t(\rho,z) \,, \quad A_\phi = A_\phi(\rho,z) \,, \quad A_z = A_\rho = 0 \,.
\end{gather}
The winding or ``vortex'' number $n_i \in \mathbb{Z}$ distinguishes between different topological solutions.

With the above ansatz, the equations of motion obtained from the action given in Eq.~\eqref{eq:action2} are
\begin{align}\label{eq:eom}
0 &= f\partial_z^2{A_t}+\frac{\partial_\rho{A_t}}{\rho}+\partial_\rho^2{A_t}-\frac{2 q^2 {A_t}}{z^2}\left({\varphi_1}^2+\varphi_2^2\right) \,, \\
\label{eq:eomaphi}
0 &= \pd_z{f}\pd_z{A_\phi}+f\partial_z^2{A_\phi}-\frac{\partial_\rho{A_\phi}}{\rho}+\partial_\rho^2{A_\phi}
+\frac{2q}{z^2}\varphi_1^2\left(n_1-qA_\phi\right)+\frac{2q}{z^2}\varphi_2^2\left(n_2-qA_\phi\right) \,, \\
\label{eq:eompsi1}
0 &= -\frac{\varphi_1}{\rho^2}\left(qA_\phi-n_1\right)^2+\frac{q^2 {A_t}^2 {\varphi_1}}{f}-
\frac{{m_1^2}{\varphi_1}}{z^2}-\frac{{\epsilon}e^{i {(n_2-n_1)} \phi  } {\varphi_2}}{z^2}
-\frac{\eta  {\varphi_1}{\varphi_2}^2}{z^2} \notag \\
&\qquad +\left(\pd_z{f}-\frac{2f}{z}\right)\pd_z{\varphi_1} + f\pd_z^2{\varphi_1}+\frac{\partial_\rho{\varphi_1}}{\rho}+\partial_\rho^2{\varphi_1} \,, \\
\label{eq:eompsi2}
0 &= -\frac{\varphi_2}{\rho^2}\left(qA_\phi-n_2\right)^2+\frac{q^2 {A_t}^2 {\varphi_2}}{f}
-\frac{{m_2^2}{\varphi_2}}{z^2}-\frac{{\epsilon}e^{i {(n_1-n_2)} \phi  } {\varphi_1}}{z^2}
-\frac{\eta  {\varphi_2}{\varphi_1}^2}{z^2} \notag \\
&\qquad +\left(\pd_z{f}-\frac{2f}{z}\right)\partial_z\varphi_2 + f\partial_z^2{\varphi_2}+\frac{\partial_\rho{\varphi_2}}{\rho}+\partial_\rho^2{\varphi_2} \,.
\end{align}

Near the boundary $z\to0$, the fields have the following asymptotic behaviors:
\begin{align}\label{eq:aphi}
\varphi_{i}(\rho,z) &\ra \varphi^{(1)}_{i}(\rho)z^{3-\Delta_{i}}+\varphi^{(2)}_{i}(\rho)z^{\Delta_{i}}
\,, \quad i = 1,\,2 \,, \\
A_\mu(\rho,z) &\ra a_\mu(\rho) + J_\mu(\rho)z \,, \quad
a_\mu = (\mu,0,0,a_\phi) \,, \quad J_\mu = (-\varrho,0,0,J_\phi) \,.
\end{align}
For the scalar fields, the AdS/CFT correspondence tells us to interpret $\varphi_i^{(1)}$ and $\varphi_i^{(2)}$ as the source and condensate respectively of the dual operator $\mathcal{O}_i$ with dimension $\Delta_i$ given by $\Delta_{i}(\Delta_{i}-3)=m^{2}_{i}$. For the gauge field, $a_\mu$ is to be interpreted as the potential in the dual CFT, while $J_\mu$ the conjugate current. In particular, $\mu$ is the chemical potential, while $\varrho$ is the charge density.

On the boundary $z = 0$, we impose the source-free conditions $\varphi_i^{(1)}\equiv0$ for the charged scalars at a fixed chemical potential, $\mu$, so that the breaking of the $U(1)$ is spontaneous if it happens. For the gauge field $a_{\phi}$, we impose either a Dirichlet or a Neumann boundary condition at the boundary depending on whether in the boundary theory, the vortices arise from a superfluid or a superconductor~\cite{Montull:2009fe,Domenech:2010nf}. For superfluid vortices we impose
\begin{equation}\label{eq:aphiz0}
A_\phi|_{z=0} = a_\phi(\rho) = \frac{1}{2}\rho^2 B \,,
\end{equation}
where $B = \partial_\rho a_\phi/\rho$ is a constant, and represents the external angular velocity of the superfluid system that is rotating~\footnote{Note that with the Dirichlet boundary condition, Eq.~\eqref{eq:aphiz0}, we may alternatively interpret the boundary theory as a superconductor in the limit where the gauge coupling is sent to zero while keeping constant the B field, which is to be thought of as an external magnetic field frozen to some constant value. We retain the superconductor notation here for this connection and also easy comparison with Refs.~\cite{Montull:2009fe,Domenech:2010nf}. We emphasize and reiterate here that when viewing the boundary theory as a superfluid as we do in the main text, $B$ is not an applied magnetic field but an external angular velocity of the rotating superfluid system.}, while for superconductor vortices we impose
\begin{equation}
\partial_{z}A_{\phi}|_{z=0} = J_{\phi}(\rho)= 0 \,.
\end{equation}
These are the boundary conditions at $z = 0$ consistent with the AdS/CFT correspondence.
At the horizon $z = z_h$, we require the fields to be regular; in particular, we require $A_t|_{z=z_h} = 0$ as usual.

We will consider a finite system with radius $R$, which we take to be much larger than the vortex radius. The boundary conditions at $\rho=R$ for superfluid vortices are given by
\begin{equation}
\partial_{\rho}\varphi|_{\rho=R}=0, \quad \partial_{\rho}A_{t}|_{\rho=R}=0, \quad A_\phi|_{\rho=R}=\frac12BR^2 \,.
\end{equation}
For superconductor vortices, the same boundary conditions apply except now $A_\phi|_{\rho=R}=n$.

Boundary conditions at $\rho=0$ are the same for both superfluid and superconductor vortices. For $n\neq0$, they are
\begin{equation}
\varphi|_{\rho=0}=0, \quad \partial_{\rho}A_{t}|_{\rho=0}=0, \quad A_{\phi}|_{\rho=0}=0 \,.
\end{equation}
For $n=0$, the boundary condition on the scalar changes to $\partial_{\rho}\varphi|_{\rho=0}=0$.

In order to avoid the divergence in energy from multiple fractional magnetic flux~\cite{Speight}, we shall set
$n_1\equiv n_2=n\in\mathbb{Z}$ henceforth.

\section{\label{sec:vortices} Superfluid/superconductor vortices}
To find the vortex solutions, we numerically solve the equations of motion (EoMs) given by Eqs.~\eqref{eq:eom}, \eqref{eq:eomaphi}, \eqref{eq:eompsi1}, and~\eqref{eq:eompsi2}, employing the pseudo-spectral Chebyshev method. For the discretization, we use a Gauss-Lobatto grid, and we set 20 grid points for the bulk $z$-direction, and 40 for the radial $\rho$-direction. After translating the EoMs as well as the boundary conditions into a system of non-linear algebraic equations, which we set up as a matrix equation using the Chebyshev differential matrices, we then solve by using the Newton-Raphson method; the error tolerance is set at $10^{-6}$. Our numerics is implemented using \textsc{matlab}.

In our numerical calculations, we work in units where $L = 1$. We set $q=1$, $n_1=n_2=n$, $m_1^2=-2, m_2^2=-5/4$, and $R=8$. We have checked that the solution obtained with $R$ extended up to 24 differ little with that obtained with $R=8$, the fractional difference being less than $10^{-4}$. We shall also set $\eta=0$ since we will not consider the effects of the density coupling here. Below we compute various properties of the superfluid vortices, varying the parameters that include the Josephson coupling, $\epsilon$, the constant external angular velocity of the rotating superfluid, $B$, and the dimensionless ratio of chemical potential to temperature, $\bar{\mu}\equiv\mu/T$~\footnote{When varying $\bar\mu$, we may think of having $\mu$ fixed while varying $T$.}.

\subsection{\label{subsec:sfvsol} Superfluid vortex solutions at various Josephson couplings}
\begin{figure}[htbp]
\includegraphics[trim=0.cm 2.5cm 0.6cm 2.5cm, clip=true,scale=0.28]{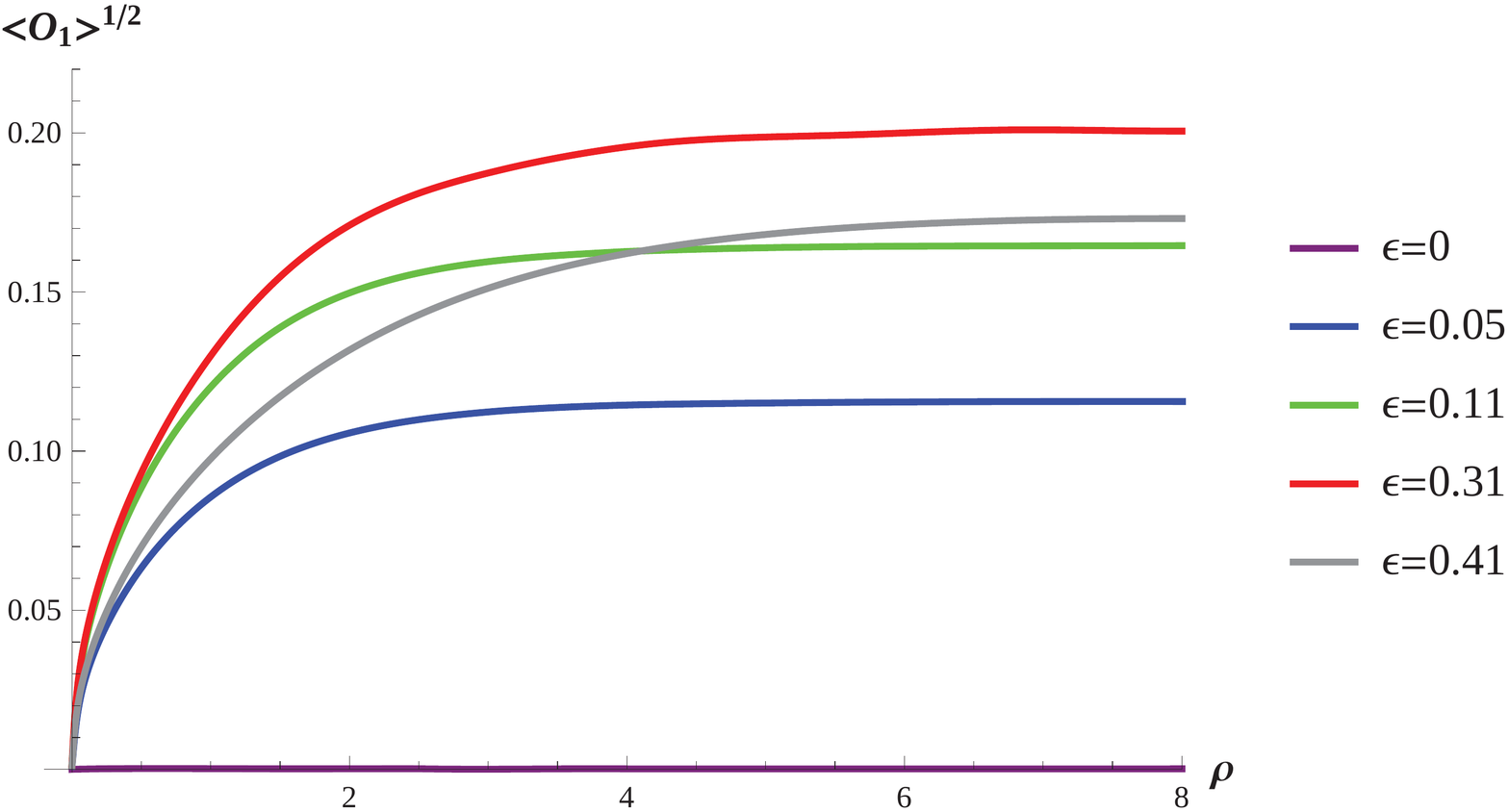}
\includegraphics[trim=0.cm 2.5cm 0.6cm 2.5cm, clip=true,scale=0.28]{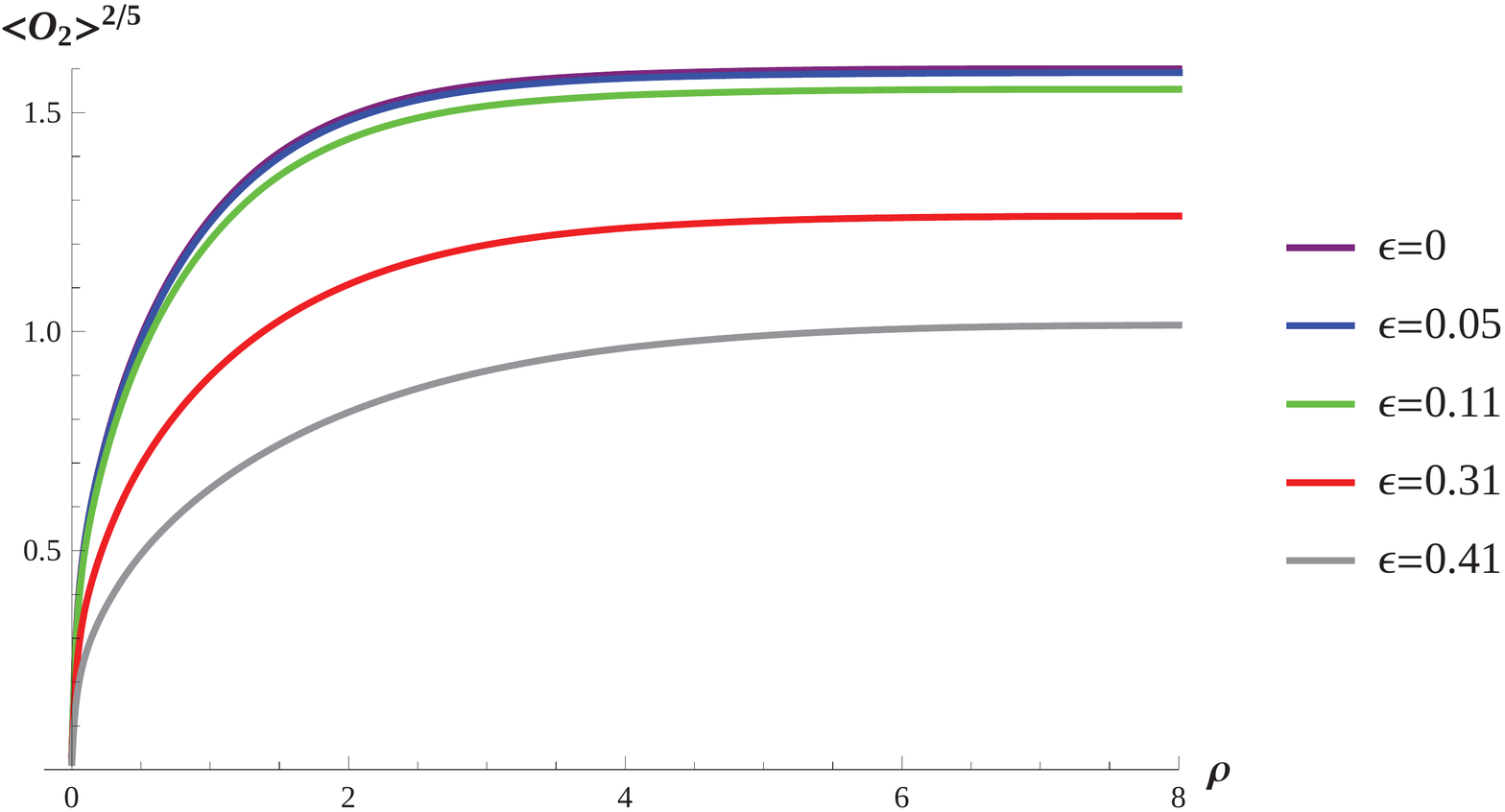}
\caption{\label{fig:O12epssmall} Order parameters $\langle O_1\rangle$ (left panel) and
$\langle O_2\rangle$ (right panel) for the $n=1$ vortex configuration at $B=0.03125$ and $\bar{\mu} = 6.2$ at various Josephson couplings $\epsilon$.}
\end{figure}
We show in Fig.~\ref{fig:O12epssmall} order parameters of a superfluid, $n=1$ single vortex solution for various values of the Josephson coupling $\eps$, with the external angular velocity set at $B = 0.03125$, and
$\bar{\mu} = 6.2$. Note that when $\eps = 0$, only the scalar $\psi_2$ condenses but not $\psi_1$, i.e.
$\langle O_1 \rangle \equiv 0$ while $\langle O_2 \rangle$ forms at a critical temperature given by $\bar\mu_c = 5.81$. It is only when $\eps$ is nonzero that $\psi_1$ also condenses, and both condensates form at the same critical temperature~\cite{Wen:2013ufa}. So we see in Fig.~\ref{fig:O12epssmall}, at $\eps = 0$ only $\langle O_2 \rangle$ is non-vanishing. But once $\eps$ is turned on, both condensates, $\langle O_{1,2} \rangle$, became nontrivial at the same critical temperature, $\bar{\mu}_c = 5.81$.

Here and below we work at $\bar{\mu} = 6.2$, which translates to $T = 0.937T_c$. We have obtained superfluid vortex solutions at other values of the temperature from just below $T_c$ to $T = 0.5T_c$. We have checked that the features we show here and below persist at other values of the temperature.

\begin{figure}[htbp]
\includegraphics[trim=0cm .5cm 0cm 0.5cm, clip=true,scale=0.24]{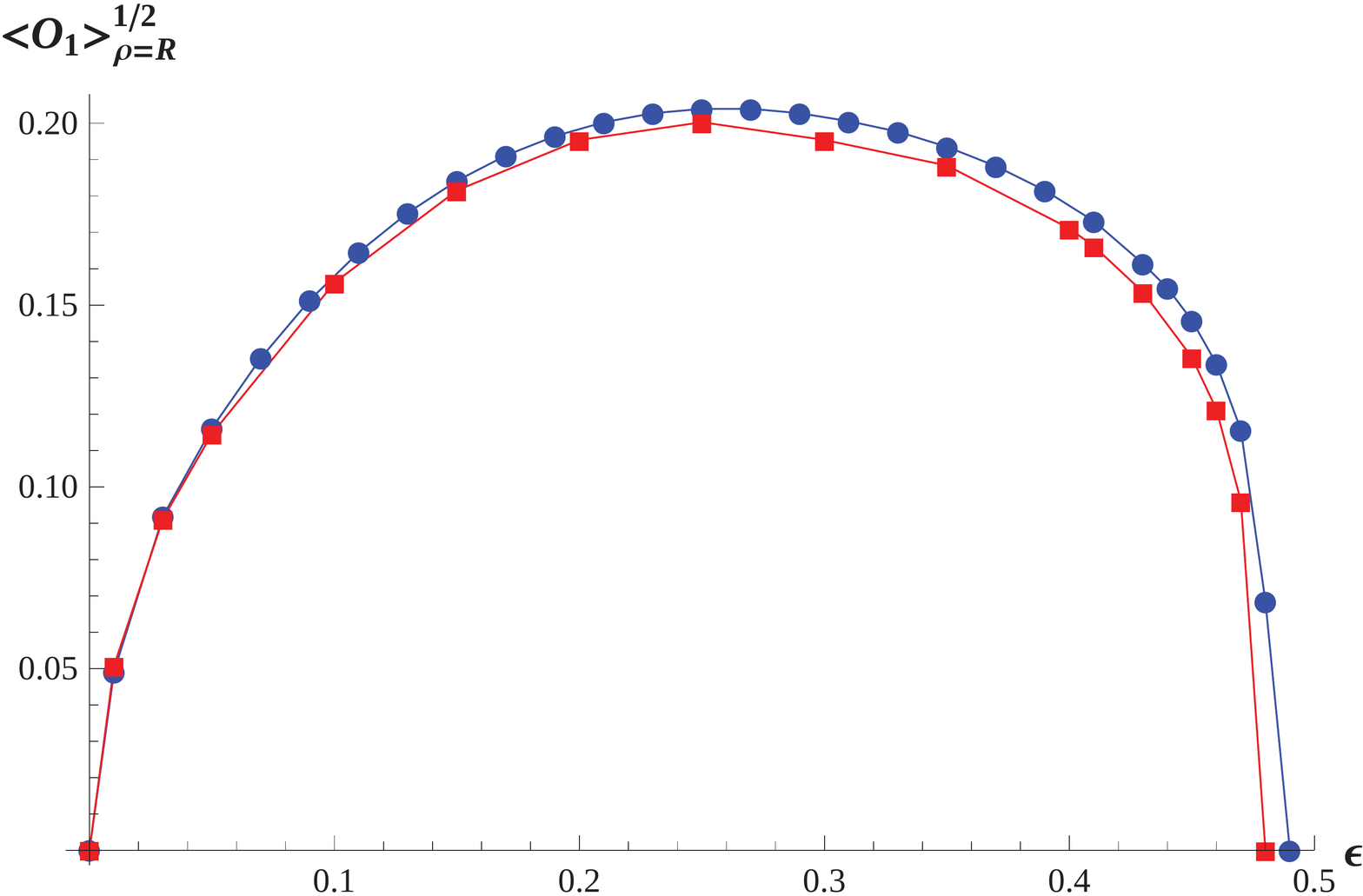}
\includegraphics[trim=0cm 2.8cm 0cm 2.8cm, clip=true,scale=0.30]{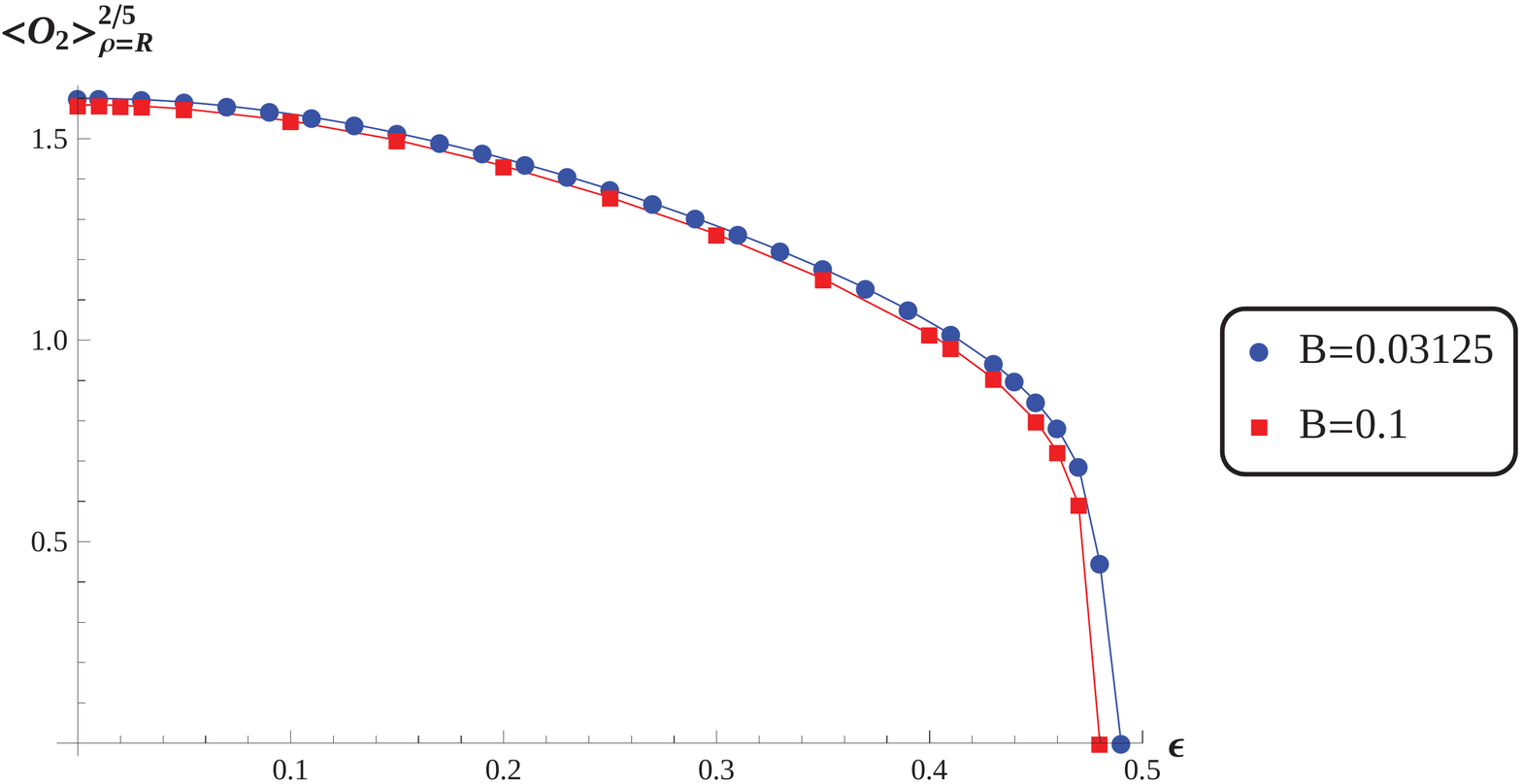}
\caption{\label{fig:O12epsbig} Value of the order parameters at the system boundary, $\langle O_1\rangle_{\rho = R}$ (left panel) and $\langle O_2\rangle_{\rho = R}$ (right panel), as functions of the Josephson coupling, $\epsilon$, for the $n=1$ vortex configuration at $B=0.03125$ and 0.1, and $\bar{\mu}=6.2$.}
\end{figure}

Fig.~\ref{fig:O12epssmall} show the usual radial behavior of the order parameter: it is zero at $\rho=0$, the vortex core, and tends to a constant far away from the core. Note that while $\langle O_2 \rangle$ seem to decrease monotonically as $\eps$ increases, $\langle O_1 \rangle$ does not behave monotonically at all. To better demonstrate the dependence of the scalar condensates on $\eps$, we plot in Fig.~\ref{fig:O12epsbig} the values of the condensates at the system boundary $\langle O_{1,2} \rangle_{\rho = R}$ as functions of $\eps$. We have plotted this $\eps$ dependence for two values of $B$, and we see that $B$ has little effect qualitatively.

From Fig.~\ref{fig:O12epsbig}, we see clearly that $\langle O_2\rangle_{\rho = R}$ decreases monotonically as $\eps$ increases; the rate of decrease is quite slow for $\eps \lesssim 0.1$. In contrast, $\langle O_2\rangle_{\rho = R}$ first increases as $\eps$ is increased from zero, turns at $\eps \approx 0.26$, and then decreases with increasing $\eps$. Note that there is a critical value, $\eps_c \approx 0.5$~\footnote{The precise value of $\eps_c$ does depend on $B$, albeit quite weakly, which is reduced as $B$ increases. For $B = 0.03125$, $\eps = 0.49$, while for $B = 0.1$, $\eps = 0.48$.}, above which both scalar condensates vanish and not only at $\rho = R$, i.e. we can no longer find a superfluid solution for a Josephson coupling above $\eps_c$, only the normal state solution with both
$\langle O_{1,2}\rangle \equiv 0$.

\subsection{Free energy and the critical angular velocity $B_{c1}$}
The free energy can be calculated holographically from the properly renormalized on-shell action. For the holographical two-band model, the (bare) on-shell action is given by
\begin{align}\label{eq:os}
S_{os} &= -\frac{1}{4\kappa^2}\int\!d^4x\partial_a\left[
\sqrt{-g}\left(A_bF^{ab}+\psi_1^*\partial^a\psi_1
+\psi_1\partial^a\psi_1^*+\psi_2^*\partial^a\psi_2+\psi_2\partial^a\psi_2^*\right)\right] \notag \\
&\quad\, +\frac{iq}{4\kappa^2}\int\!d^4x\sqrt{g}A_b\left[
\psi_1^*\left(\partial^b-iqA^b\right)\psi_1-\psi_1\left(\partial^b+iqA^b\right)\psi_1^*
+ (\psi_1\leftrightarrow\psi_2)\right]  \notag \\
&\quad\, +\frac{\eta}{2\kappa^2}\int\!d^4x\sqrt{-g}|\psi_1|^2|\psi_2|^2 \,.
\end{align}
Note that terms involving $\epsilon$ have been removed by the equations of motion.

The first term in Eq.~\eqref{eq:os} produces a surface integral. To remove the divergence coming from it, we need to add the counterterm
\begin{equation}
S_{ct}=\frac{-1}{2\kappa^2}\int d^3x \sqrt{-\gamma}\left(\psi_1\psi_1^*\right)+\frac{-1/2}{2\kappa^2}\int d^3x\sqrt{-\gamma}\left(\psi_2\psi_2^*\right) \,,
\end{equation}
where $\gamma$ is a reduced metric on the boundary with $\sqrt{-\gamma}=\rho/z^3$. Adding all the contributions together, we obtain the free energy from the finite, regularized on-shell action
\begin{align}
F &= -TS_{reg.} = -T\left(S_{os}+S_{ct}\right) \notag \\
&= \frac{-T}{2\kappa^2}\int\!dtd\phi\,\bigg\{
\int\!d\rho\rho\left(\frac{\varrho\mu}{2}+\frac{BJ_\phi}{4}\right)\bigg|_{{z=0}}
-\int\!dz\frac{A_\phi\partial_\rho A_\phi}{2\rho}\bigg|_{\rho=R}
\notag \\
&\qquad +\int\!dzd\rho\,\frac{\rho}{z^2}\left[-\frac{q^2 A_t^2(\varphi_1^2+\varphi_2^2)}{f}
-\frac{qA_\phi}{\rho^2}\left[\varphi_1^2\left(n_1-qA_\phi\right)+\varphi_2^2(n_2-qA_\phi)\right]
+\frac{\eta \varphi_1^2\varphi_2^2}{z^2}\right]\bigg\} \,.
\end{align}
Note that we have kept the density coupling, $\eta$, for completeness above. Since we do not consider its effect here, it is set to zero below in our numerical calculations.

\begin{figure}[htbp]
\includegraphics[trim=0cm 1.0cm 0cm 1.0cm, clip=true,scale=0.26]{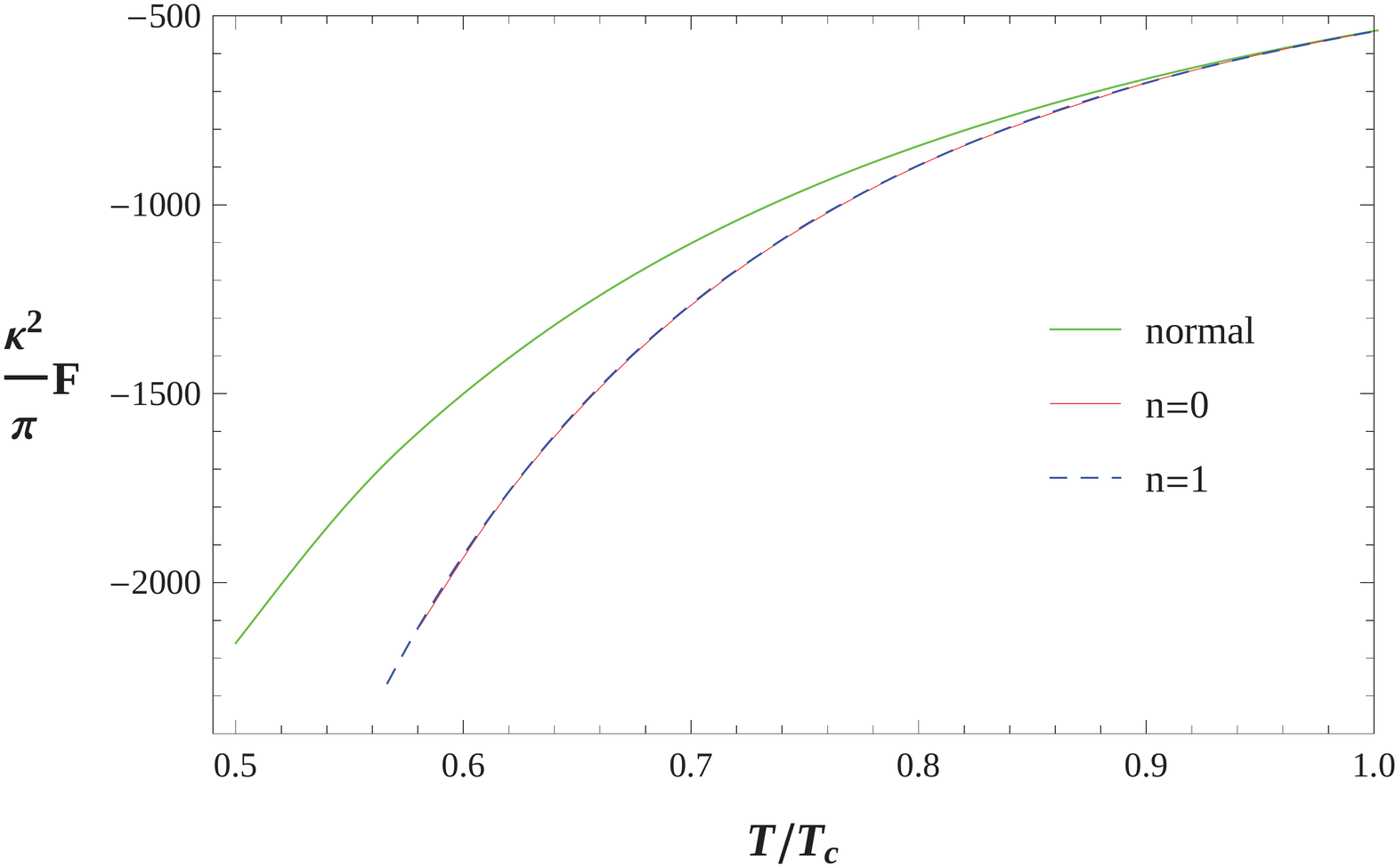}
\includegraphics[trim=0cm 1.0cm 0cm 1.0cm, clip=true,scale=0.26]{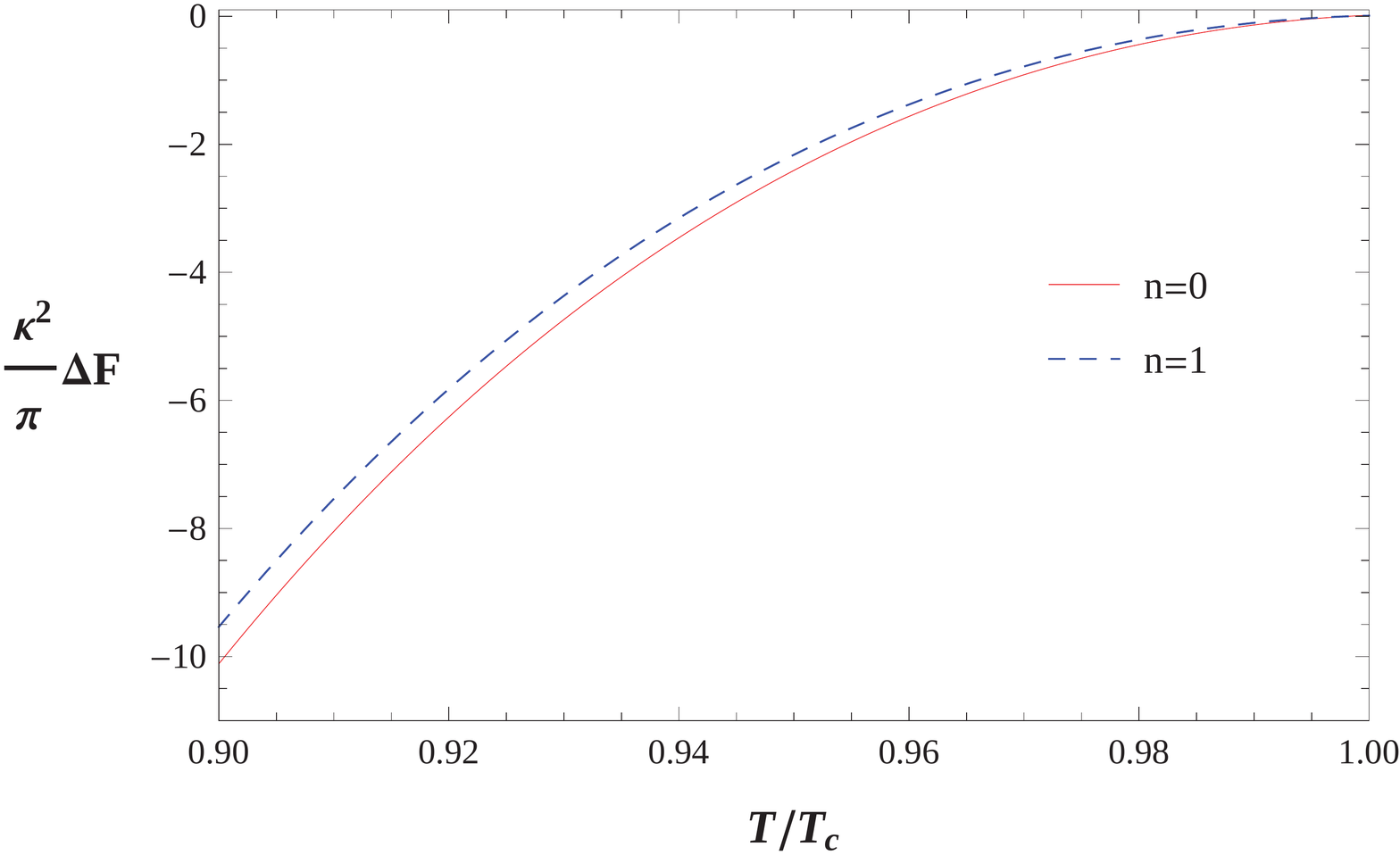}
\caption{\label{fig:freeT} (Left) Free energy of the normal state solution, and the superfluid, $n=0$ and $n=1$ vortex solutions as a function of temperature. The green line is the free energy for the normal state solution, the red (blue dashed) line the $n=0$ ($n=1$) vortex solution. (Right) Free energy difference of the $n=0$ and $n=1$ vortex solution to the normal state solution. In all cases, the solutions are obtained at $\epsilon=0.05$ and $B=0.03125$.}
\end{figure}
We show in Fig.~\ref{fig:freeT} the temperature dependence of the free energy for the normal state
(non-superfluid) solution, and the superfluid, $n=0$ and $n=1$ vortex solutions at $\epsilon=0.05$ and $B=0.03125$. As already mentioned above, the critical temperature at which the scalar condensate forms is given by $\bar\mu_c = 5.81$.

From the left panel of Fig.~\ref{fig:freeT}, we see that below $T_c$ when superfluid forms, the superfluid solutions have lower free energy than the normal state solution, as expected of the superfluid being thermodynamically favored below $T_c$. Next, to distinguish which is thermodynamically favored, the $n=0$ or the $n=1$ vortex solution, we plot in the right panel of Fig.~\ref{fig:freeT} for each winding configuration the free energy difference between the superfluid vortex solution and the normal state solution
\begin{equation}
\Delta F = F(\varphi_i \neq 0)-F(\varphi_i=0) \,,
\end{equation}
where $F(\varphi_i=0)$ ($F(\varphi_i \neq 0)$) denotes the free energy of the normal state (superfluid vortex) solution with both scalars vanishing (condensing). We see that for $B=0.03125$ and $\epsilon=0.05$, the $n=0$ is preferred over the $n=1$ vortex solution. Note that we have displayed only the region close to $T_c$ so that the two solution curves can be clearly distinguished.

\begin{figure}[htbp]
\includegraphics[trim=0cm 2.8cm 0cm 2cm, clip=true,scale=0.28]{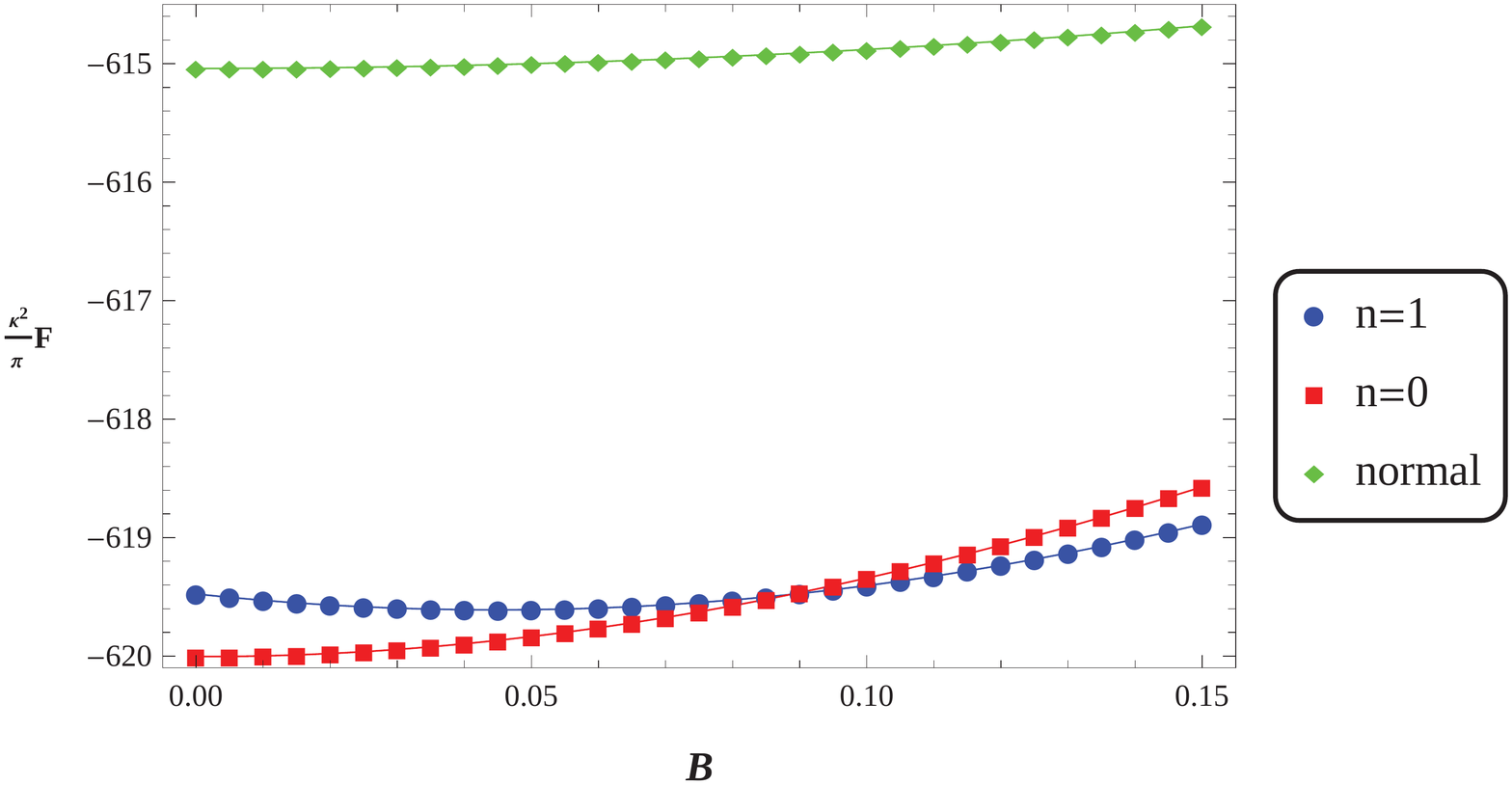}
\includegraphics[trim=0cm 2.5cm 0cm 2cm, clip=true,scale=0.27]{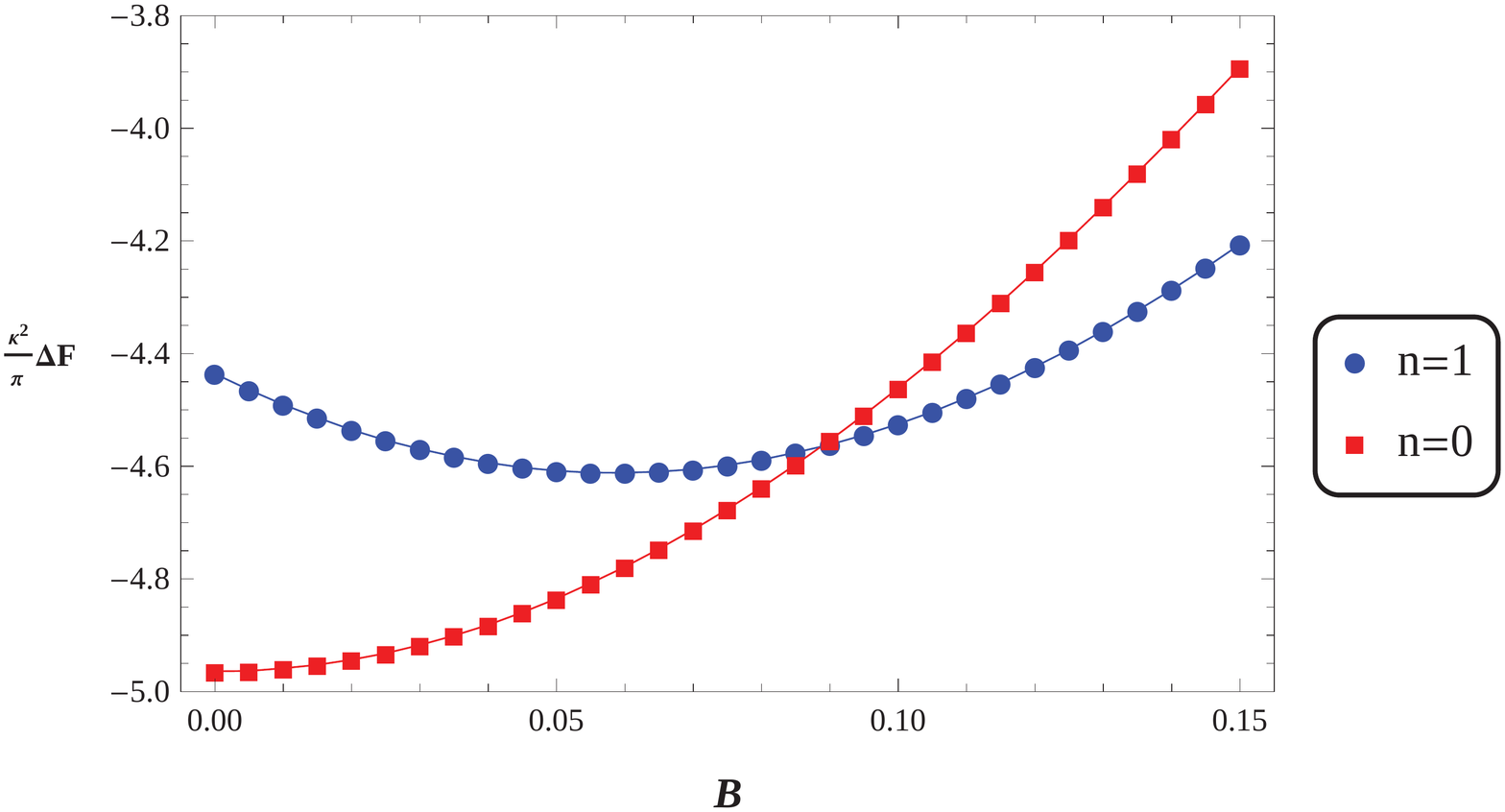}
\caption{\label{fig:Bc1} (Left) Free energy of the normal state and superfluid vortex solutions as a function of $B$. The green diamond, red square, and blue circle denote respectively the normal state, the $n = 0$ and $n = 1$ vortex solutions. (Right) Free energy difference of the $n=0$ and $n=1$ vortex solution to the normal state solution. In all cases, the solutions are obtained at $\epsilon=0.05$ and $T = 0.937T_c$.}
\end{figure}
We show in Fig.~\ref{fig:Bc1} the $B$ dependence of the free energy of the normal state and the superfluid vortex solutions at $\epsilon=0.05$ and $T = 0.937T_c$. We see that there is a critical value, $B_{c1} = 0.09$, where the free energy for both the $n = 1$ and the $n = 0$ configurations coincide, and so marks the beginning for which the
$n = 1$ vortex solution becomes thermodynamically favored over the $n = 0$ one.

\begin{figure}[htbp]
\includegraphics[trim=0cm 0cm 0cm 0cm, clip=true,scale=0.26]{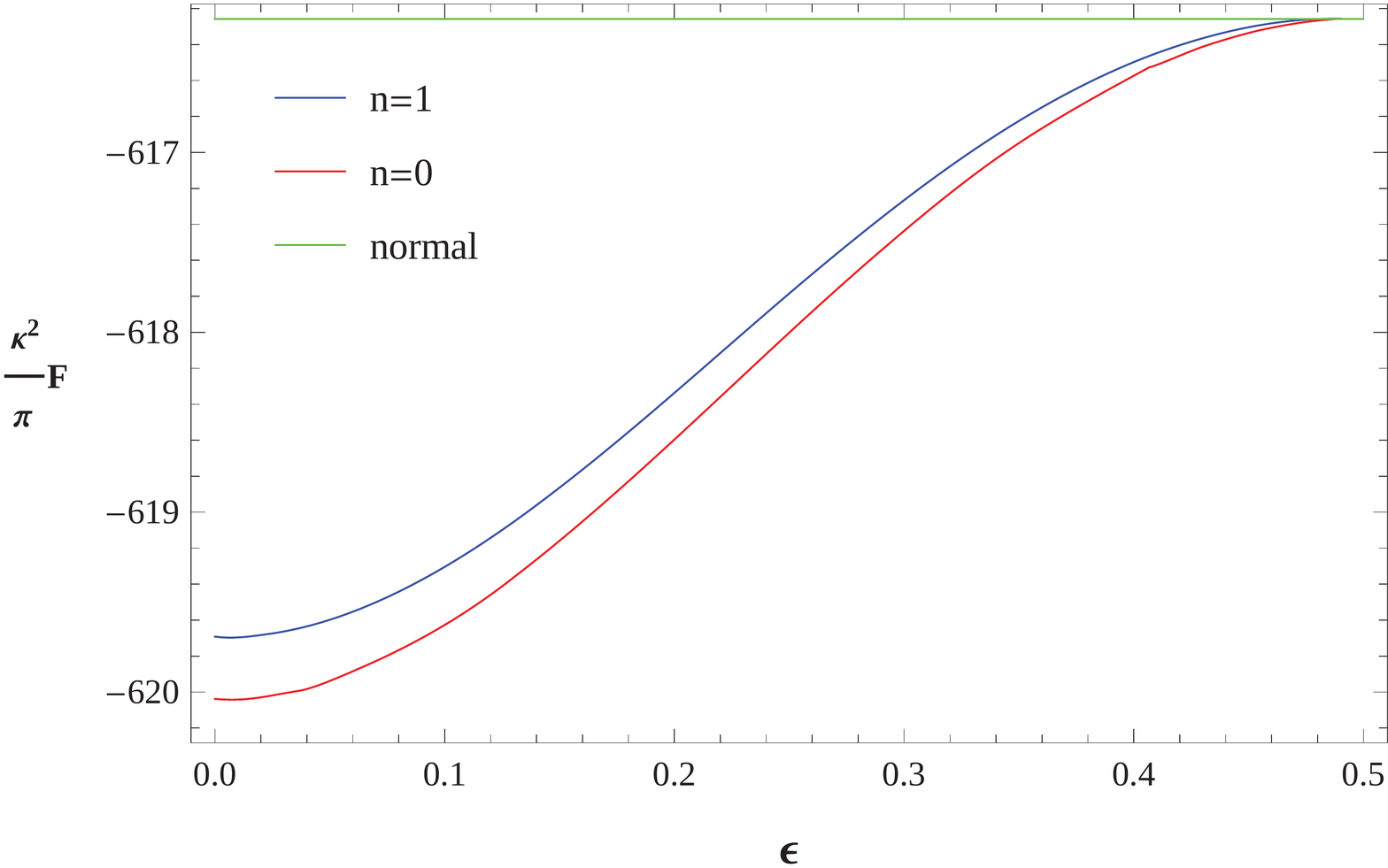}
\includegraphics[trim=0cm 0cm 0cm 0cm, clip=true,scale=0.26]{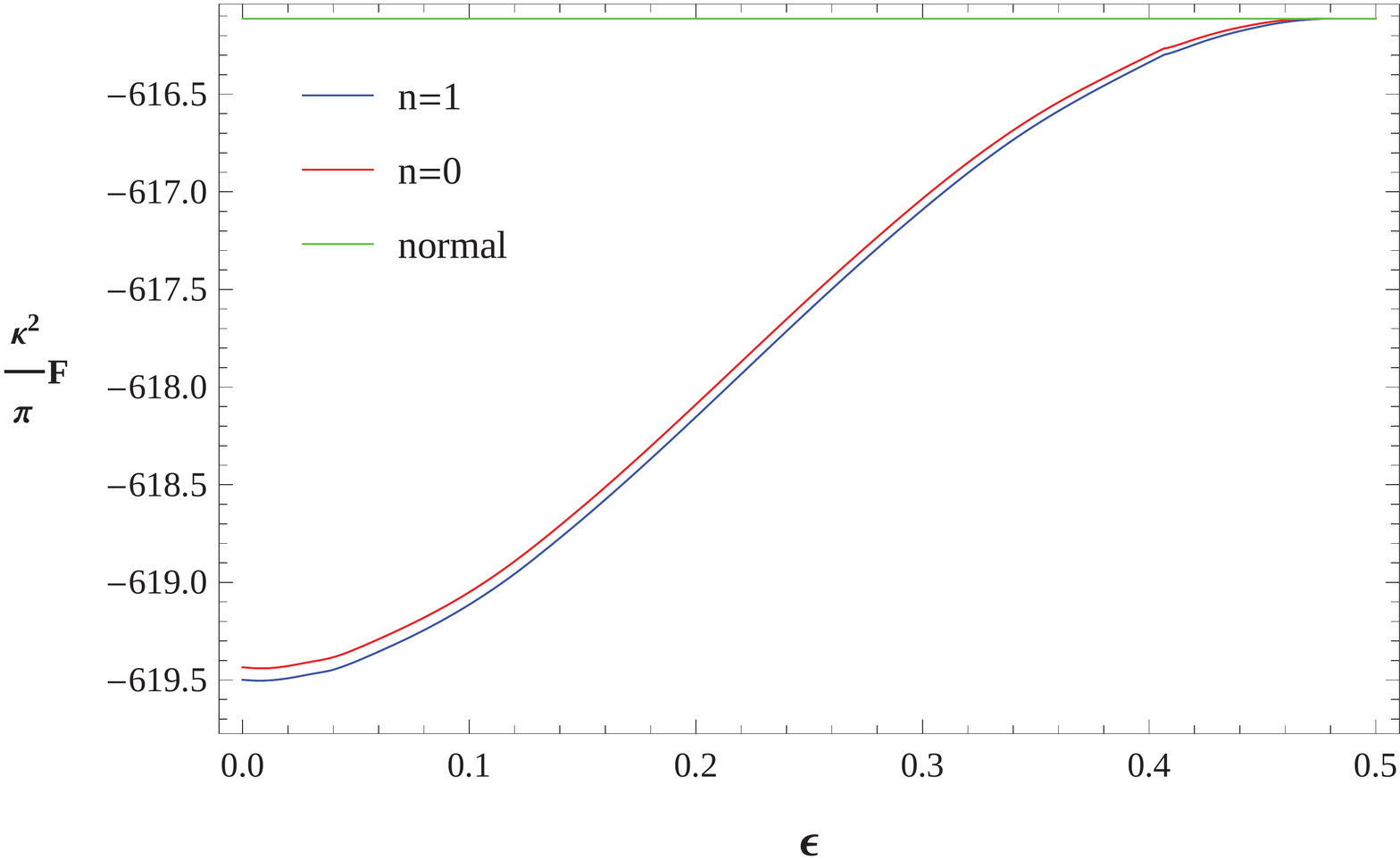}
\includegraphics[trim=3.7cm 3.7cm 3.7cm 3.7cm, clip=true,scale=0.85]{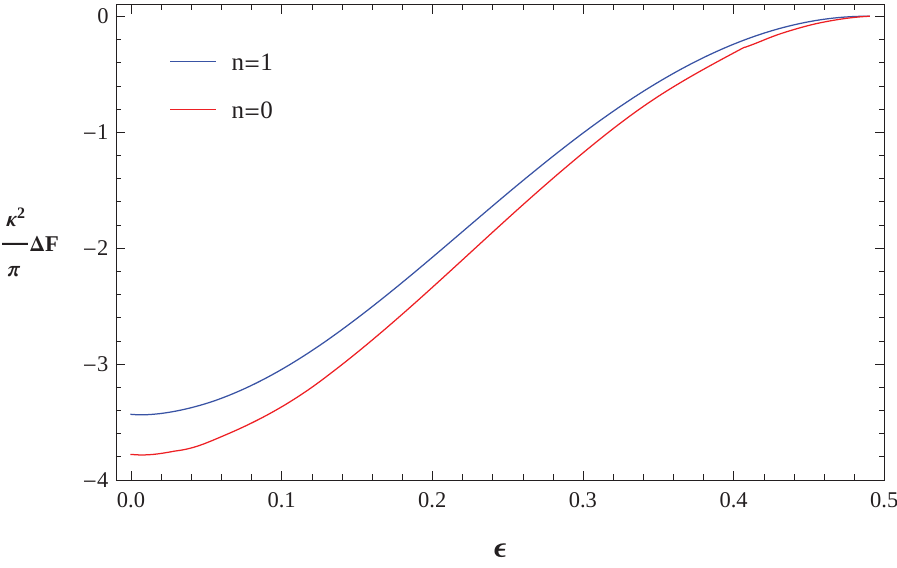}
\includegraphics[trim=3.7cm 3.7cm 3.7cm 3.7cm, clip=true,scale=0.85]{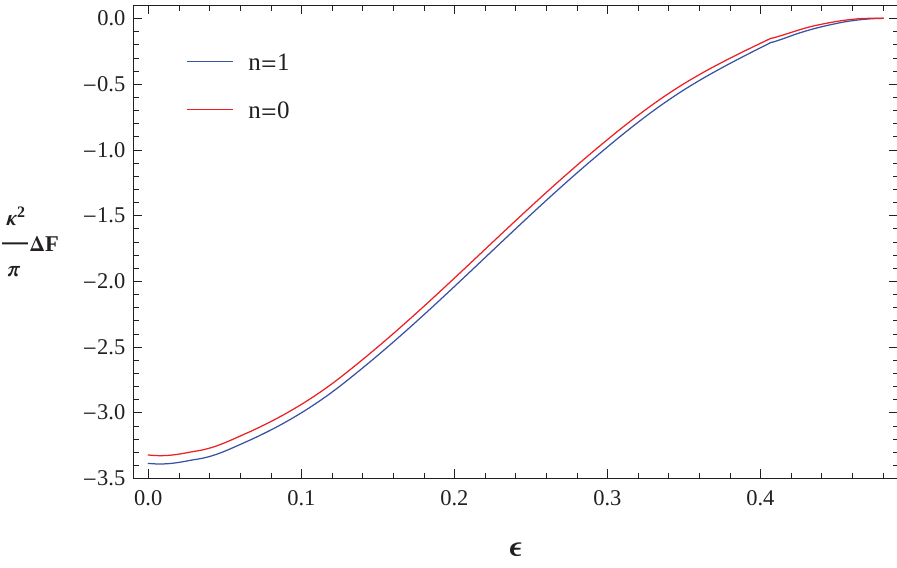}
\caption{\label{fig:Fdiff} Free energy (top row) and free energy difference (bottom row) as a function of Josephson coupling, $\eps$, for $B=0.03125 < B_{c1}$ (left column) and $B = 0.1 > B_{c1}$ (right column). In all cases, the solutions are obtained at $T = 0.937T_c$.}
\end{figure}
We show in Fig.~\ref{fig:Fdiff} the dependence of the free energy of the normal state and the superfluid vortex solutions on the Josephson coupling for both $B < B_{c1}$ and $B > B_{c1}$ at $T = 0.937T_c$. We see that for the range of $\eps$ shown, when $B < B_{c1}$, the $n = 0$ solution is favored over the $n = 1$ one, and so has the lower free energy and thus larger $|\Delta F|$, while when $B < B_{c1}$, the reverse is true. Next, we see that when $\eps$ approaches $\eps_c(B) \approx 0.5$ for the values of $B$ used here, the free energy of both the $n = 0$ and $n = 1$ vortex solutions approach that of the normal state solution. This reflects the fact that above $\eps_c(B)$ no superfluid solution can be found in our numerics, only the normal state solution. This feature was already seen in Fig.~\ref{fig:O12epsbig}.

\subsection{Superfluid density and coherence lengths}
By the AdS/CFT correspondence, the superfluid density, $n_s$, can be obtained from the conjugate current, $J_\phi$, as~\cite{Montull:2009fe}
\begin{equation}
n_s=\frac{J_\phi}{n-a_\phi} \,,
\end{equation}
where $a_\phi = \half\rho^2 B$ with $B$ the external angular velocity. Note that the denominator $n-a_\phi$ is the gauge-invariant superfluid velocity along the angular direction,
$v_\phi=(\nabla\arg[\psi_i])_\phi-a_\phi$~\cite{Tinkham}.

\begin{figure}[htbp]
\includegraphics[trim=0cm 1.0cm 0cm .8cm, clip=true,scale=0.26]{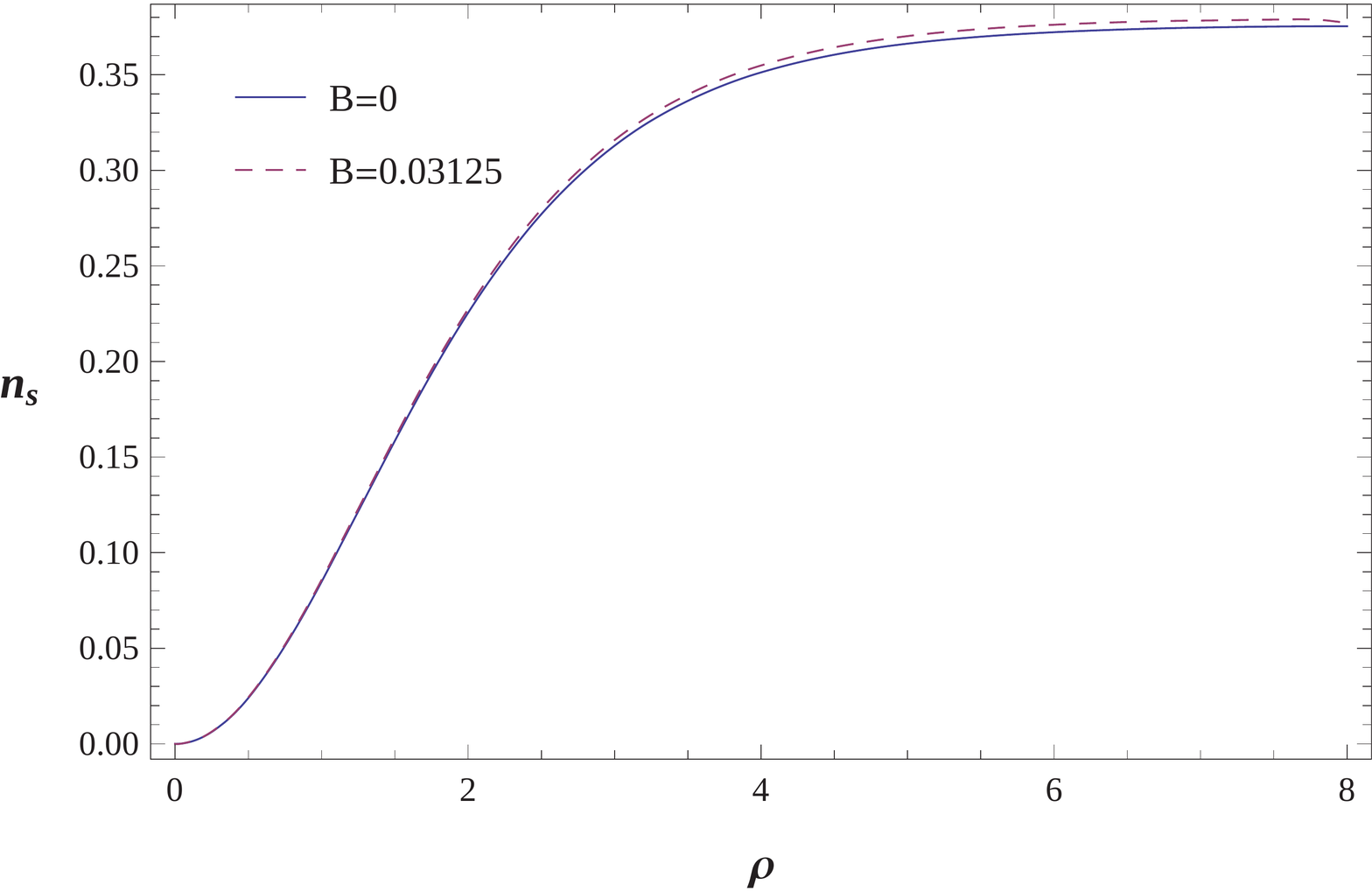}
\includegraphics[trim=0cm 1.0cm 0cm .8cm, clip=true,scale=0.26]{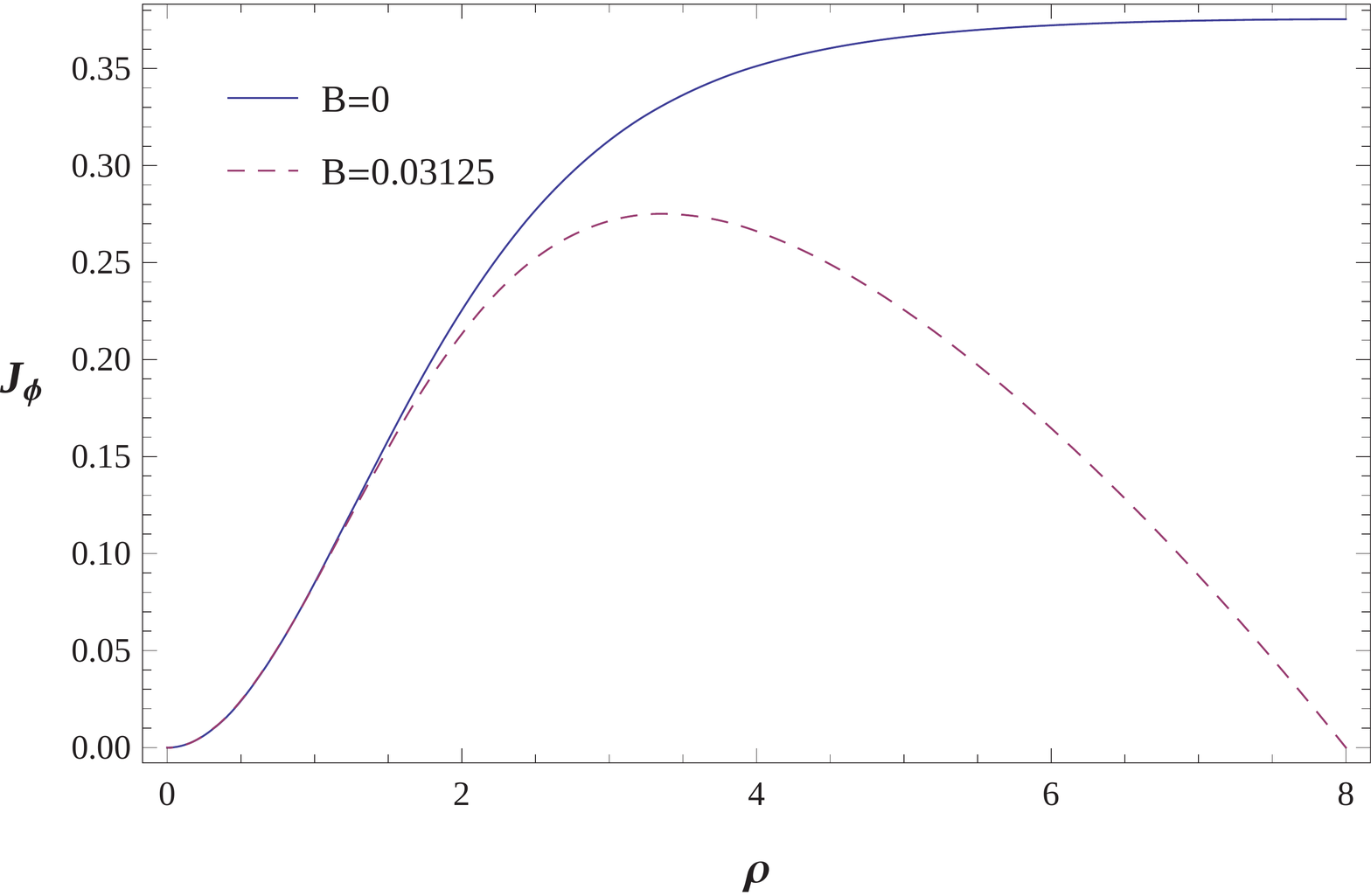}
\caption{\label{fig:ns} Superfluid density $n_s$ and current $J_\phi$ for the $n = 1$ vortex configuration with $B=0$ (solid lines) and $B=0.03125$ (dashed lines) at $T = 0.937T_c$ and $\epsilon=0.05$.}
\end{figure}
We show in Fig.~\ref{fig:ns} the profile of $n_s$ and $J_{\phi}$ in the radial $\rho$-direction for the
$n = 1$ configuration at $B=0$ and 0.03125. For $n_s$, we see that external rotation has a little effect on the superfluid density. But for $J_\phi$, when there is external rotation ($B \neq 0$), after rising from zero at the vortex core, instead of approaching a nonzero, finite constant far away from the core, $J_\phi$ drops back to zero at some distance from the core. This reflects the fact that $n_s$ stays finite and nonzero whether there is external rotation or not, but the superfluid angular velocity $v_\phi = n - a_\phi = n - \half\rho^2 B$ will become zero at some $\rho > 0$ when there is external rotation.

\begin{figure}[htbp]
\includegraphics[trim=0cm 2.1cm 0cm 2.0cm, clip=true,scale=0.27]{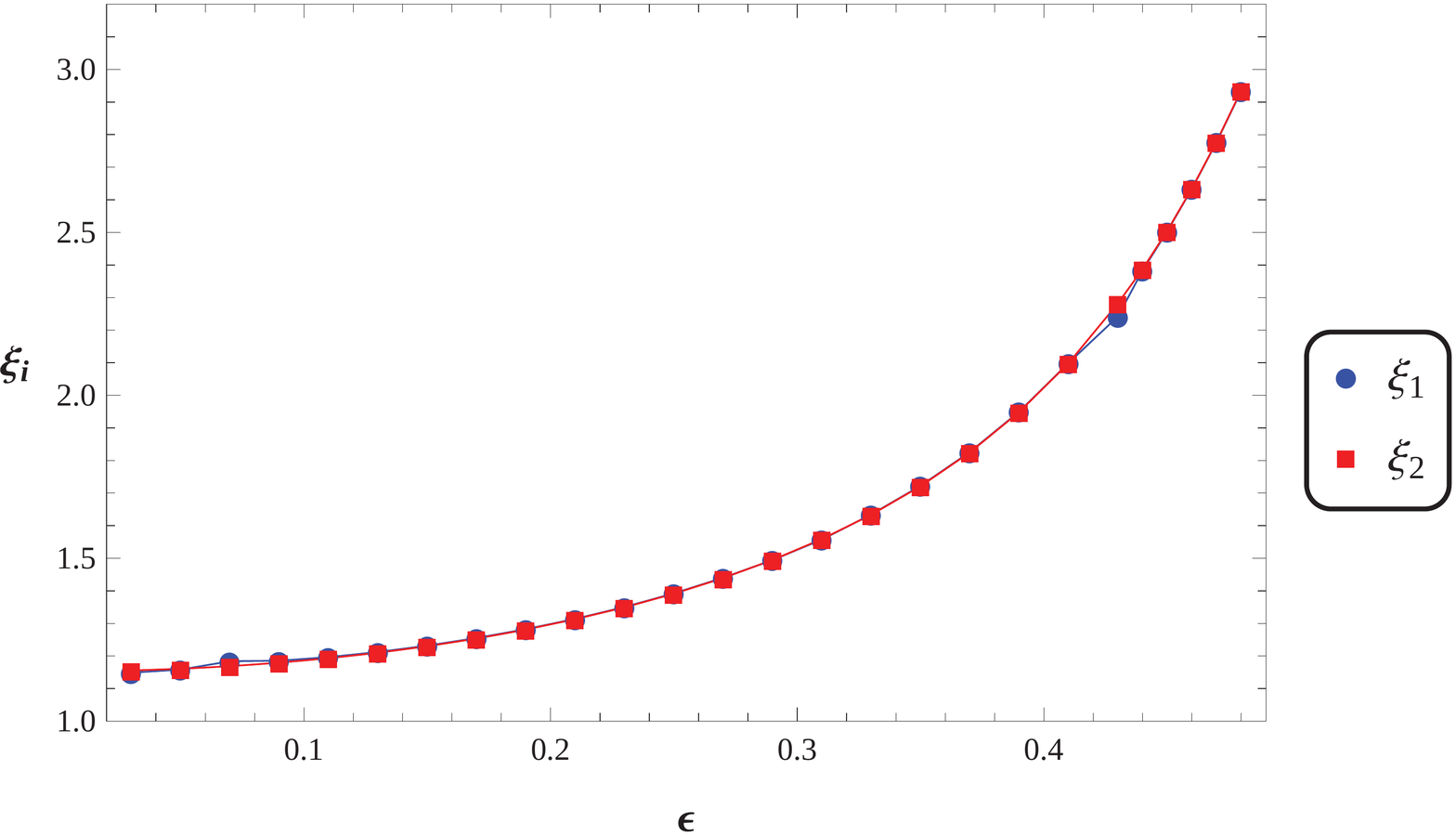}
\includegraphics[trim=0cm 1.8cm 0cm 1.95cm, clip=true,scale=0.27]{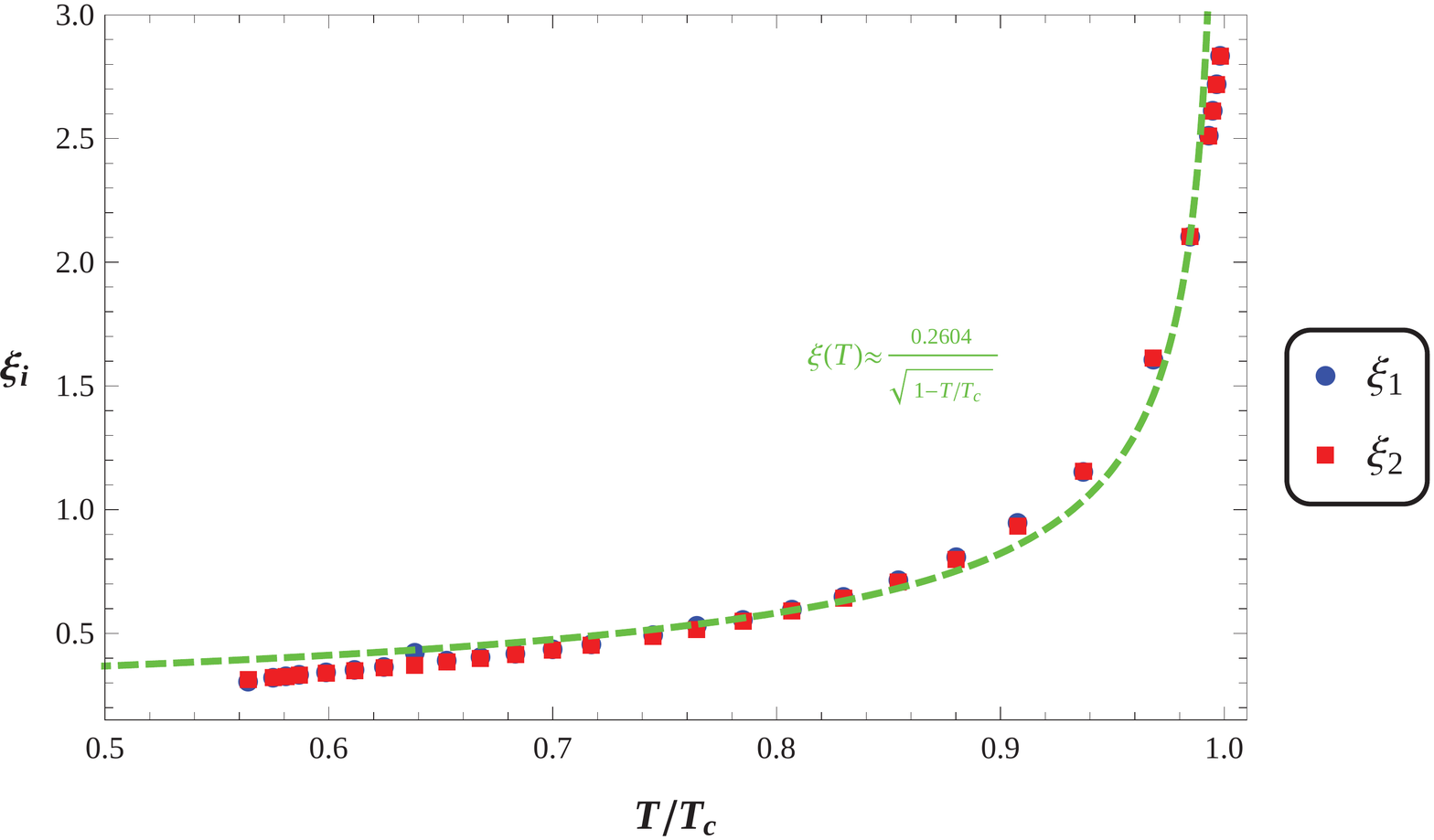}
\caption{\label{fig:lamxiT} Dependence of the coherence lengths on $\epsilon$ at $T = 0.937T_c$ (left panel) and temperature at $\eps = 0.05$ (right panel) for the $n = 1$ vortex configuration with $B=0.03125$. On the right panel, the dashed green line is a fit to the temperature dependence.}
\end{figure}

For a two-band superfluid, we expect there to be two condensates circulating around the vortex core, and thus two coherence lengths, $\xi_i$, corresponding to each condensate. The coherence length can be extracted from the condensate itself~\cite{annett}:
\begin{equation}\label{eq:annett}
\langle O_i(\rho)\rangle =O_i(\infty)\tanh\left(\frac{\rho}{\sqrt2\xi_i}\right) \,,
\end{equation}
where $O_i(\infty)$ denotes the asymptotic value of the condensate. In Fig~\ref{fig:lamxiT}, we show the dependence of the coherence lengths on $\eps$ and the temperature for the $n = 1$ vortex configuration.
We see that coherence lengths increases as both $\eps$ and temperature increases, and as $T$ approaches $T_{c}$, the coherence length diverges as it should. We see also that the two coherence lengths are very close to each other throughout the range of $\eps$ we looked at, whether for small $\eps \ll 0.1$ or for $\eps$ close to $\eps_c$. We have checked that these features persist for other values of $B$, both above and below $B_{c1}$.

In Fig~\ref{fig:lamxiT}, close to $T_c$ the coherence lengths have the form $\xi_i(T) = 0.2604(1-T/T_c)^{-1/2}$, which is the expected temperature dependence from the GL theory. Another feature we see immediately is that the two coherence lengths differ very little from each other (barring numerical errors). Close to $T_c$, this is expected from the GL theory. But it is surprising to find that this behavior persists down to low temperatures. A possible reason for this may be that the Josephson coupling is locking the growth and the saturation of the condensates together. We will investigate the mechanism behind this in future works.

\subsection{Superconductor vortex}
We consider now superconductor vortices. In this case, the external magnetic field $B$ is dynamical, and we can thus see the screening of $B$.
\begin{figure}[htbp]
\includegraphics[trim=0cm 2.55cm 0cm 2.4cm, clip=true,scale=0.28]{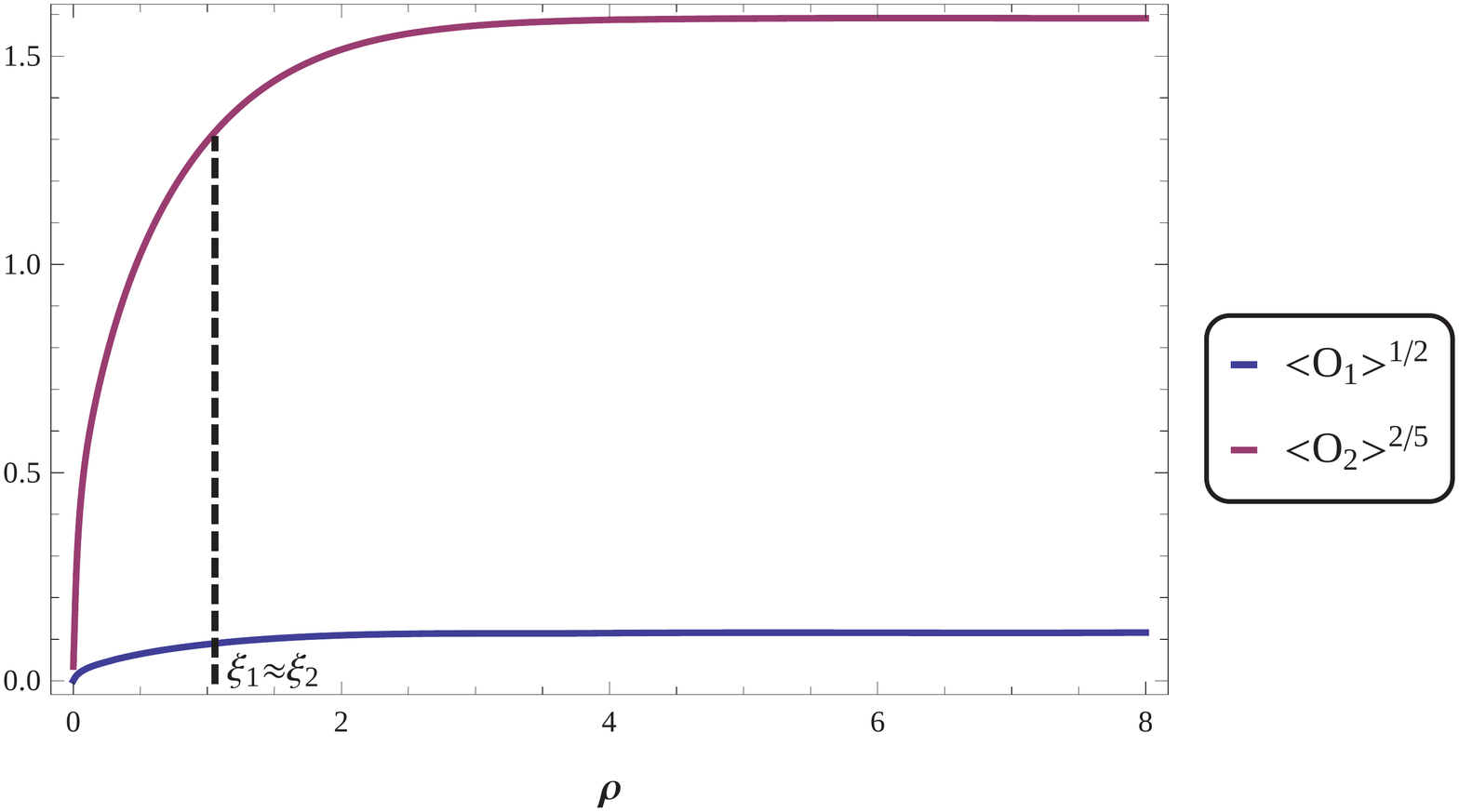}
\includegraphics[trim=0cm 2.2cm 0cm 2.2cm, clip=true,scale=0.27]{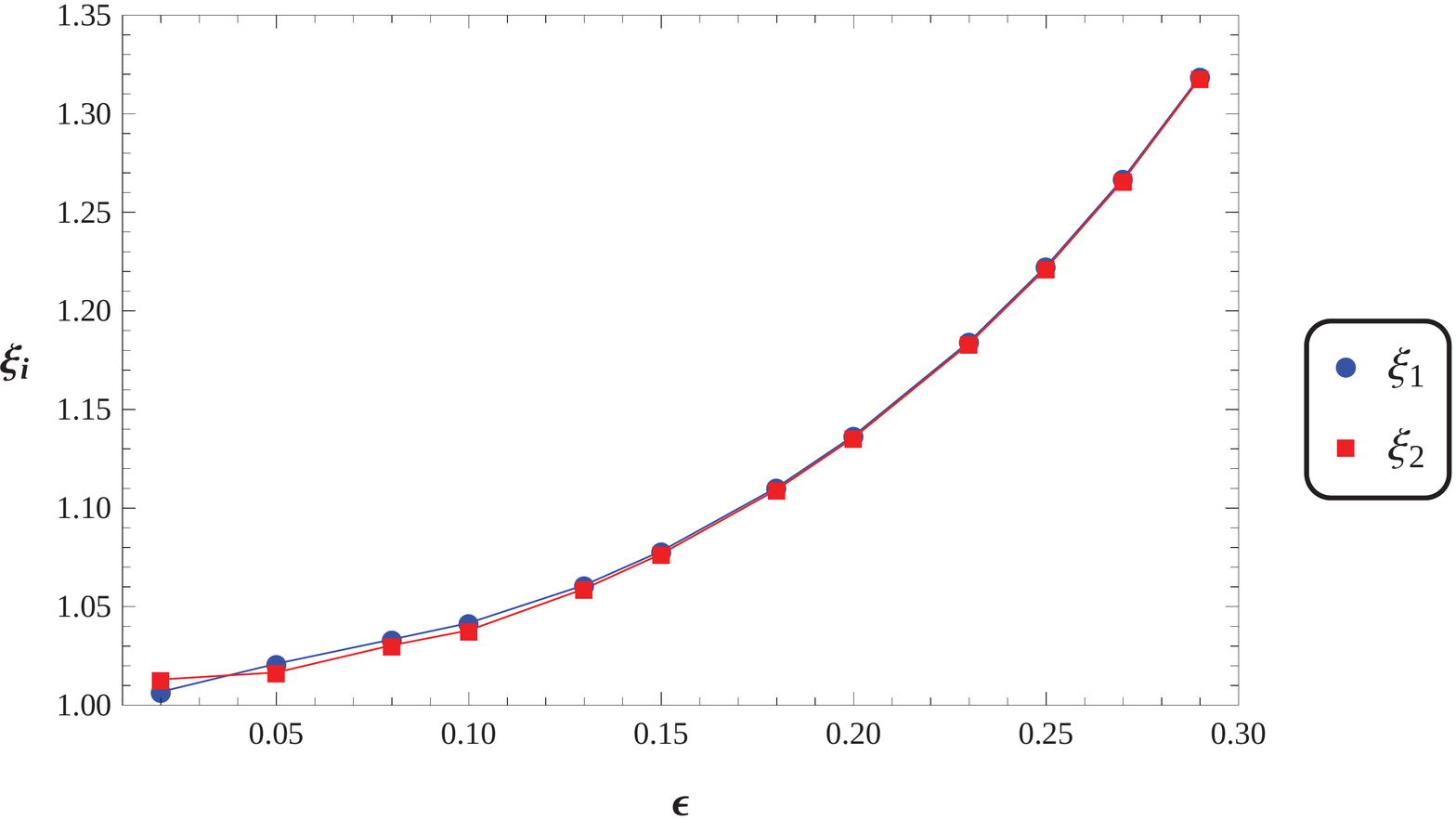}
\caption{\label{fig:xineumann} (Left panel) Superconductor order parameters for the $n=1$ vortex configuration at
$T=0.937T_c$ and $\epsilon=0.05$. The dashed line marks the coherence lengths, $\xi_{i}$. (Right panel) Coherence lengths, $\xi_i$, of the superconductor vortex as a function of the Josephson coupling, $\epsilon$.}
\end{figure}

\begin{figure}[htbp]
\includegraphics[trim=0cm 0.9cm 0cm 0.7cm, clip=true,scale=0.3]{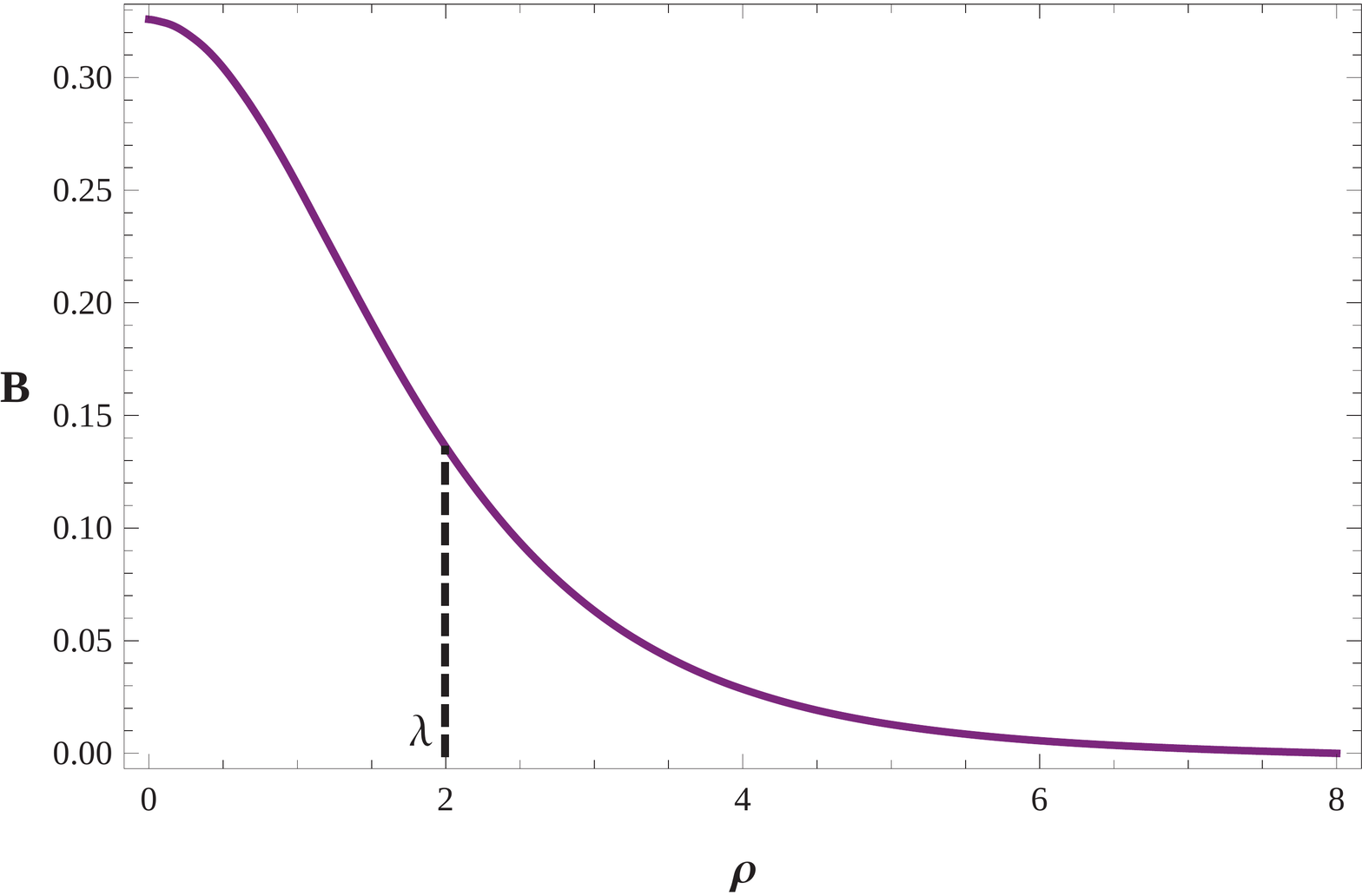}
\caption{\label{fig:lamneumann} Profile of the magnetic field $B$ at $T=0.937T_c$ and $\epsilon=0.05$. The dashed line marks the penetration depth $\lambda$.}
\end{figure}
We show in the left panel of Fig.~\ref{fig:xineumann} the profile of the superconducting order parameters (scaled to have unit mass dimension) inside the superconductor for the $n=1$ configuration at $T=0.937T_c$ and $\epsilon=0.05$. The coherence lengths, $\xi_i$, can be extracted as in the superfluid case using the form given in Eq.~\eqref{eq:annett}, and we show in the right panel of Fig.~\ref{fig:xineumann} their dependence on $\eps$. We see that the two superconductor coherence lengths stay very close to each other throughout the range of $\eps$ we looked at.

We show in Fig.~\ref{fig:lamneumann} the profile of the magnetic field $B$ inside the superconductor for the $n = 1$ configuration at $T=0.937T_c$ and $\epsilon=0.05$. The magnetic penetration length can be extracted from
$B =  b e^{-\rho/\lambda}$. At $T=0.937T_c$ and $\epsilon=0.05$, we obtain $\xi_1=1.02106$, $\xi_2=1.01662$, and $\lambda=2.06235$. Calculating the GL parameters, $\kappa_{i}=\lambda/\xi_{i}$, we get $\kappa_1=2.01981$ and $\kappa_2=2.02862$, which are within $0.4\%$ to each other. Note that $\kappa_{1,2} > 1/\sqrt{2}$, which indicates that we have a type II superconductor.

\begin{figure}[htbp]
\includegraphics[trim=0cm 1.8cm 0cm 1.8cm, clip=true,scale=0.27]{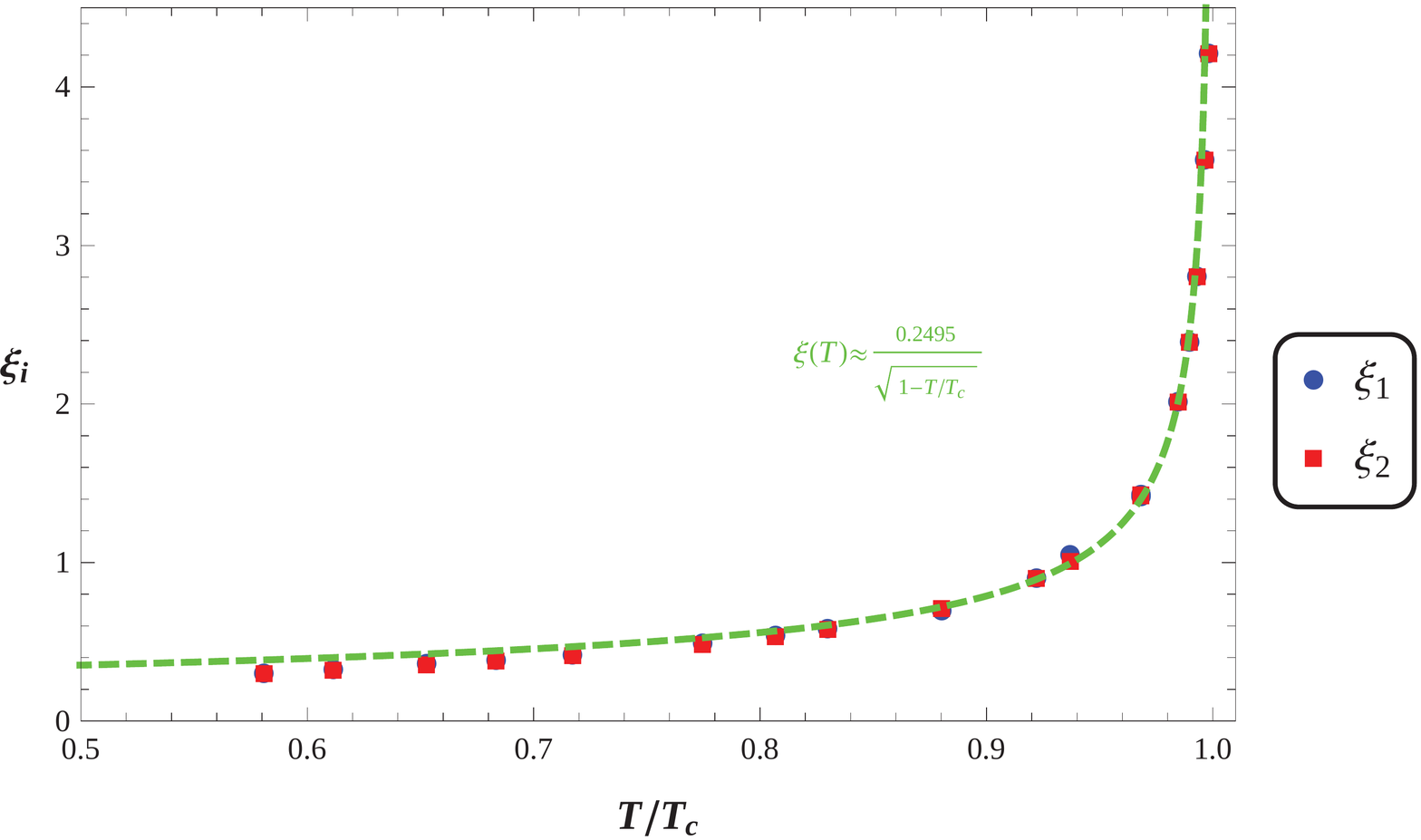}
\includegraphics[trim=0cm 0.5cm 0cm 0.45cm, clip=true,scale=0.23]{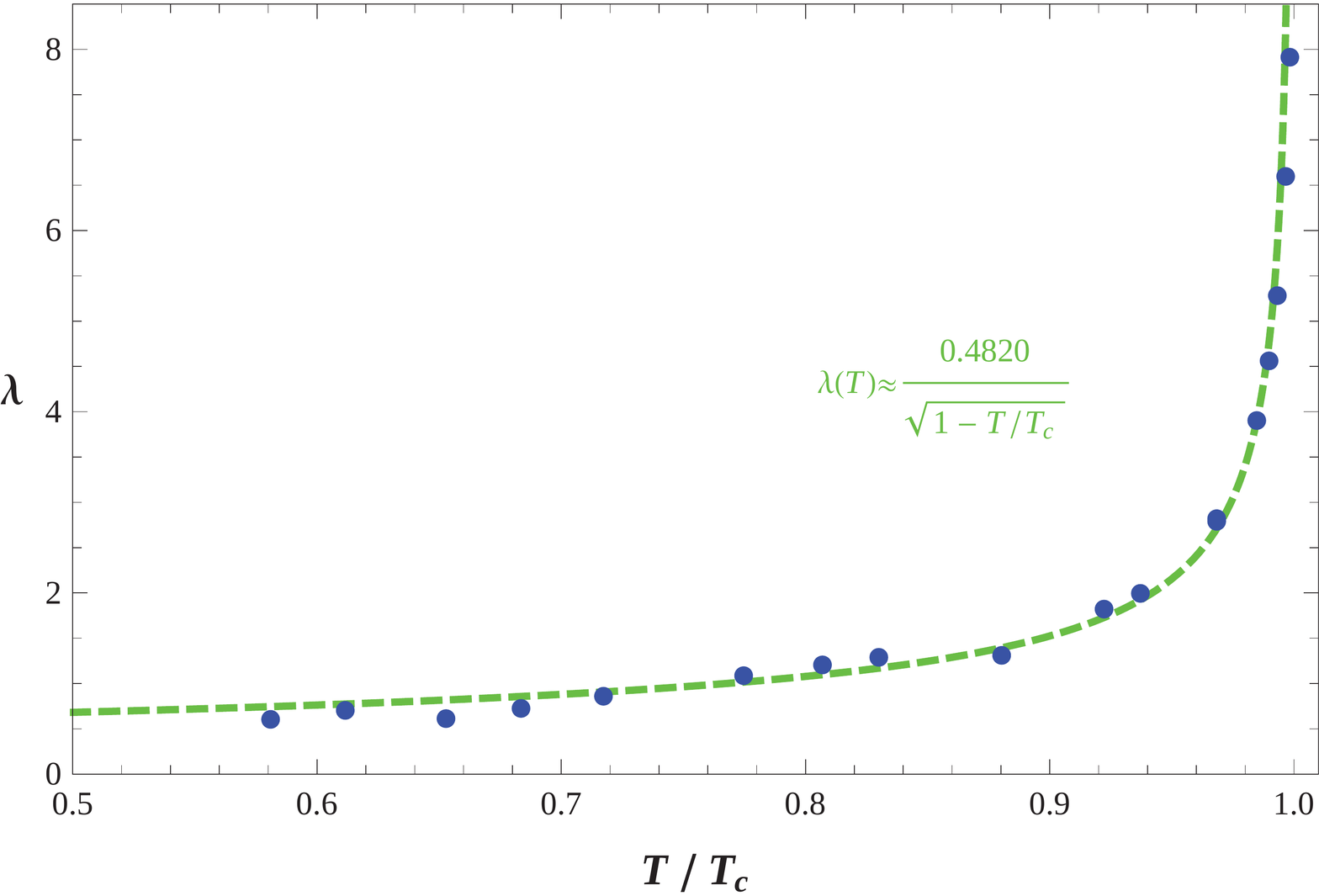}
\caption{\label{fig:xilamTneumann} Temperature dependence of coherence lengths $\xi_i$ (left panel) and penetration length $\lambda$ (right panel) for the $n=1$ vortex configuration at $\epsilon=0.05$. The dashed green lines shows the  the fitted temperature dependence.}
\end{figure}

We show in Fig.~\ref{fig:xilamTneumann} the temperature dependence of $\xi_i$ and $\lambda$. Near $T_c$, we have good fits from $\xi(T) = 0.2495/\sqrt{1-T/T_c}$ and $\lambda(T) = 0.4820/\sqrt{1-T/T_c}$. We see that there are very little difference between $\xi_1$ and $\xi_2$ down to $T \sim 0.5 T_{c}$. Computing the GL parameter $\kappa_i$ for temperature range considered here, we find $\kappa_{1,2} > 1/\sqrt{2}$ over the entire range, indicating a type II superconductor down to $T \sim 0.5 T_{c}$.

\section{\label{sec:concl} Summary and Outlook}

In this paper, we have studied the magnetic response of a holographic two-band superconductor that has an interband Josephson coupling between the two bulk complex scalars. We have constructed the single vortex solution and study the effects of the Josephson coupling. By imposing appropriate boundary conditions, we can consider both superfluid and superconductor vortices. For superfluid vortices, we find one condensate is insensitive to the Josephson coupling when it is below 0.1. By comparing the free energy of $n = 0$ and $n = 1$ vortex configurations, we have estimated the first critical magnetic field. We have also extracted coherence lengths from the condensates for both the superfluid and superconductor cases, as well as the magnetic penetration length in the superconductor case where the magnetic field is dynamical, and we can see explicit screening. Near the critical temperature, we have checked that the temperature dependence of the coherence lengths are consistent with GL theory. Surprisingly, for
both the superfluid and superconductor vortices we find there is effective only one coherence length in the range of parameters we consider,
leading to the virtually the same GL parameter for both bands. Furthermore, the GL parameters are all greater than $1/\sqrt{2}$ for the whole temperature range that our numerics is reliable, indicating that our holographic two-band superconductor is type II, and the absence of type-1.5 superconductivity.

The paper is a fist step in the study of vortex dynamics in strongly-coupled/correlated multiband superconductors employing holography. There are many interesting future directions to take. An immediate one is to scan over a larger parameter space by going to larger Josephson coupling and different bulk scalar masses. Another would be to go beyond the static case studied here and construct dynamical vortex solutions. This would allow us to study interactions between vortices at different distances, and would allow a direct check on the dynamical mechanism of the purported type-1.5 superconductivity. It would also be very useful to generalize to a three-band model. There one can study
the existence of chiral and time-reversal symmetry breaking state, interband phase difference induced domain walls, fractional quantum flux vortices~\cite{three-band, three-band review} and frustrated superconductors~\cite{frustrated SC}. Lastly, it would be interesting to clarify issues surrounding hidden criticality~\cite{hidden criticality} using a holographic model of multiband superconductivity.

\section*{Acknowledgement}
We would like to thank Egor Babaev, Sean Hartnoll, Zhao Huang, Nabil Iqbal, Miroad Milosevic, Wei-Feng Tasi and Takashi Yanagisawa for discussions and comments. SYW is grateful to the response from the participants in the conference Superstripe 2015. HQZ is especially grateful to KITPC's hospitality and partial support during completing this work. MSW thanks the support of National Center for Theoretical Sciences in Taiwan during the course of this work. The work of SYW was supported by the Ministry of Science and Technology (grant no. MOST-101-2112-M-009-005 and MOST 104-2811-M-009-068) and National Center for Theoretical Sciences in Taiwan. HQZ was supported in part by the fund of Utrecht University budget associated to Gerard 't Hooft and the Young Scientists Fund of the National Natural Science Foundation of China (No.11205097).


\begin{thebibliography}{999}
\bibitem{MgB2}
J. Nagamatsu, N. Nakagawa, T. Muranaka, Y. Zenitani and J. Akimitsu, ''Superconductivity at 39K in magnesium diboride'', Nature 410 63-4 (2001).

\bibitem{iron}
Yoichi Kamihara, Takumi Watanabe, Masahiro Hirano and Hideo Hosono, ``Iron-Based Layered Superconductor
$La[O_{1-x}F_{x}]FeAs$ (x =0.05-0.12) with Tc=26K", J. Am. Chem. Soc., 2008, 130 (11), pp 3296¡V3297.

\bibitem{MgB2 review}
C.-Buzea and T. Yamashita, ``Review of superconducting properties of MgB$_{2}$ Superconductors'', Science and Technology, Vol. 14, No. 11 (2001) R115-R146, [arXiv:cond-mat/0108265 [cond-mat.supr-con]].

\bibitem{FeSC review}
P. J. Hirschfeld, M. M. Korshunov, I. I. Mazin, ``Gap symmetry and structure of Fe-based superconductors'', Rep. Prog. Phys. 74, 124508 (2011), [arXiv:1106.3712 [cond-mat.suprcon]],
David C. Johnston, ``The Puzzle of High Temperature Superconductivity in Layered Iron Pnictides and Chalcogenides'', Advances in Physics 59, 803-1061 (2010), [arXiv:1005.4392 [cond-mat.supr-con]],
G. R. Stewart, ``Superconductivity in Iron Compounds'', Rev. Mod. Phys. 83 1589-1652 (2011), [arXiv:1106.1618 [cond-mat.supr-con]].
\bibitem{TCGL}
A. A. Shanenko, M. V. Milo\v{s}evi¡¦c, F. M. Peeters and A. V. Vagov, "Extended Ginzburg-Landau Formalism for Two-Band Superconductors" Phys. Rev. Lett. {\bf 106}, 047005 (2011),[arXiv:1101.0971 [cond-mat.supr-con]].
Phys. Rev. B 86, 144514 (2012) and Phys. Rev. B 87, 134510 (2013)
M. Silaev and E. Babaev, "Microscopic derivation of two-component Ginzburg-Landau model and conditions of its applicability in two-band systems" Phys. Rev. B {\bf 85}, 134514 (2012), [arXiv:1110.1593 [cond-mat]].
\bibitem{fractional flux quanta}
Y. Tanaka, D. D. Shivagan, A. Crisan, A. Iyo, P. M. Shirage, K. Tokiwa, T. Watanabe, and N.Terada, ``Vortex molecule, fractional flux quanta, and interband phase difference soliton in multi-band superconductivity and multi-component superconductivity'', Journal of Physics: Conference Series 150 (2009) 052267.
R. Geurts, M. V. Milo\v{s}evi\.{c}, F. M. Peeters, "Vortex matter in mesoscopic two-gap superconducting disks: influence of Josephson and magnetic coupling", Phys. Rev. B {\bf 81}, 214514 (2010), [arXiv:1005.2921 [cond-mat.supr-con]].
Juan C. Pina, Clecio C. de Souza Silva, Milorad V. Milo\v{s}evi\.{c}, "Stability of fractional vortex states in a two-band mesoscopic superconductor
", Phys. Rev. B {\bf 86}, 024512 (2012), [arXiv:1205.2022 [cond-mat.supr-con]].

\bibitem{type 1.5}
Egor Babaev and Martin Speight, ``Semi-Meissner state and neither type-I nor type-II superconductivity in multicomponent superconductors'', Phys. Rev. B {\bf 72}, 180502(R) (2005), [arXiv:cond-mat/0411681 [cond-mat.supr-con]],
V. V. Moshchalkov, M. Menghini, T. Nishio, Q.H. Chen, A.V. Silhanek, V.H. Dao, L.F. Chibotaru, N. D. Zhigadlo, J. Karpinsky, ``Type 1.5 Superconductor'', Phys. Rev. Lett. {\bf 102}, 117001 (2009), [arXiv:0902.0997[cond-mat.supr-con]],
Johan Carlstrom, Egor Babaev, Martin Speight, ``Type-1.5 superconductivity in multiband systems: the effects of interband couplings'', Phys. Rev. B {\bf 83}, 174509 (2011), [arXiv:1009.2196 [cond-mat.supr-con]],
Mihail Silaev and Egor Babaev, ``Microscopic theory of type-1.5 superconductivity in multi-band systems'', Phys. Rev. B
{\bf 84}, 094515 (2011), [arXiv:1102.5734 [cond-mat.supr-con]].


\bibitem{Maldacena:1997re}
  J.~M.~Maldacena,
  ``The Large N limit of superconformal field theories and supergravity,''
  Int.\ J.\ Theor.\ Phys.\  {\bf 38}, 1113 (1999)
  [Adv.\ Theor.\ Math.\ Phys.\  {\bf 2}, 231 (1998)]
  [hep-th/9711200].
\bibitem{Witten:1998qj}
  E.~Witten,
  ``Anti-de Sitter space and holography,''
  Adv.\ Theor.\ Math.\ Phys.\  {\bf 2}, 253 (1998)
  [hep-th/9802150].
\bibitem{QCD}
  J.~Erlich, E.~Katz, D.~T.~Son and M.~A.~Stephanov,
  ``QCD and a holographic model of hadrons,''
  Phys.\ Rev.\ Lett.\  {\bf 95}, 261602 (2005)
  [hep-ph/0501128].
  A.~Karch, E.~Katz, D.~T.~Son and M.~A.~Stephanov,
  ``Linear confinement and AdS/QCD,''
  Phys.\ Rev.\ D {\bf 74}, 015005 (2006)
  [hep-ph/0602229].
  M.~Kruczenski, D.~Mateos, R.~C.~Myers and D.~J.~Winters,
  ``Towards a holographic dual of large N(c) QCD,''
  JHEP {\bf 0405}, 041 (2004)
  [hep-th/0311270].
  T.~Sakai and S.~Sugimoto,
  ``Low energy hadron physics in holographic QCD,''
  Prog.\ Theor.\ Phys.\  {\bf 113}, 843 (2005)
  [hep-th/0412141].
  S.~He, S.~Y.~Wu, Y.~Yang and P.~H.~Yuan,
  ``Phase Structure in a Dynamical Soft-Wall Holographic QCD Model,''
  JHEP {\bf 1304}, 093 (2013) [arXiv:1301.0385 [hep-th]].

\bibitem{heavy ion collision}
  S.~Caron-Huot, P.~Kovtun, G.~D.~Moore, A.~Starinets and L.~G.~Yaffe,
  ``Photon and dilepton production in supersymmetric Yang-Mills plasma,''
  JHEP {\bf 0612}, 015 (2006)
  [hep-th/0607237].
  S.~Y.~Wu and D.~L.~Yang,
  ``Holographic Photon Production with Magnetic Field in Anisotropic Plasmas,''
  JHEP {\bf 1308}, 032 (2013)
  [arXiv:1305.5509 [hep-th]].
  B.~Muller, S.~Y.~Wu and D.~L.~Yang,
  ``Elliptic flow from thermal photons with magnetic field in holography,''
  Phys.\ Rev.\ D {\bf 89}, no. 2, 026013 (2014)
  [arXiv:1308.6568 [hep-th]].
  A.~Rebhan, A.~Schmitt and S.~A.~Stricker,
  ``Anomalies and the chiral magnetic effect in the Sakai-Sugimoto model,''
  JHEP {\bf 1001}, 026 (2010)
  [arXiv:0909.4782 [hep-th]].
  S.~Pu, S.~Y.~Wu and D.~L.~Yang,
  ``Holographic Chiral Electric Separation Effect,''
  Phys.\ Rev.\ D {\bf 89}, no. 8, 085024 (2014)
  [arXiv:1401.6972 [hep-th]].
  S.~Pu, S.~Y.~Wu and D.~L.~Yang,
  ``Chiral Hall Effect and Chiral Electric Waves,''
  Phys.\ Rev.\ D {\bf 91}, no. 2, 025011 (2015)
  [arXiv:1407.3168 [hep-th]].

  \bibitem{Hartnoll:2008vx}
  S.~A.~Hartnoll, C.~P.~Herzog and G.~T.~Horowitz,
  ``Building a Holographic Superconductor,''
  Phys.\ Rev.\ Lett.\  {\bf 101}, 031601 (2008)
  [arXiv:0803.3295 [hep-th]].

\bibitem{Hartnoll:2008kx}
  S.~A.~Hartnoll, C.~P.~Herzog and G.~T.~Horowitz,
  ``Holographic Superconductors,''
  JHEP {\bf 0812}, 015 (2008)
  [arXiv:0810.1563 [hep-th]].

\bibitem{Gubser:2008wv}
  S.~S.~Gubser and S.~S.~Pufu,
  ``The Gravity dual of a p-wave superconductor,''
  JHEP {\bf 0811}, 033 (2008)
  [arXiv:0805.2960 [hep-th]].

\bibitem{Chen:2010mk}
  J.~W.~Chen, Y.~J.~Kao, D.~Maity, W.~Y.~Wen and C.~P.~Yeh,
  ``Towards A Holographic Model of D-Wave Superconductors,''
  Phys.\ Rev.\ D {\bf 81}, 106008 (2010)
  [arXiv:1003.2991 [hep-th]].

\bibitem{Benini:2010pr}
  F.~Benini, C.~P.~Herzog, R.~Rahman and A.~Yarom,
  ``Gauge gravity duality for d-wave superconductors: prospects and challenges,''
  JHEP {\bf 1011}, 137 (2010)
  [arXiv:1007.1981 [hep-th]].

\bibitem{Kim:2013oba}
  K.~Y.~Kim and M.~Taylor,
  ``Holographic d-wave superconductors,''
  JHEP {\bf 1308}, 112 (2013)
  [arXiv:1304.6729 [hep-th]].

\bibitem{Wen:2013ufa}
  W.~Y.~Wen, M.~S.~Wu and S.~Y.~Wu,
  ``Holographic model of a two-band superconductor,''
  Phys.\ Rev.\ D {\bf 89}, no. 6, 066005 (2014)
  [arXiv:1309.0488 [hep-th]].
\bibitem{Huang:2011ac}
  C.~Y.~Huang, F.~L.~Lin and D.~Maity,
  ``Holographic Multi-Band Superconductor,''
  Phys.\ Lett.\ B {\bf 703}, 633 (2011)
  [arXiv:1102.0977 [hep-th]].
\bibitem{Krikun:2012yj}
  A.~Krikun, V.~P.~Kirilin and A.~V.~Sadofyev,
  ``Holographic model of the $S^{\pm}$ multiband superconductor,''
  JHEP {\bf 1307}, 136 (2013)
  [arXiv:1210.6074 [hep-th]].
\bibitem{Montull:2009fe}
  M.~Montull, A.~Pomarol and P.~J.~Silva,
  ``The Holographic Superconductor Vortex,''
  Phys.\ Rev.\ Lett.\  {\bf 103}, 091601 (2009)
  [arXiv:0906.2396 [hep-th]].

\bibitem{Domenech:2010nf}
  O.~Domenech, M.~Montull, A.~Pomarol, A.~Salvio and P.~J.~Silva,
  ``Emergent Gauge Fields in Holographic Superconductors,''
  JHEP {\bf 1008}, 033 (2010)
  [arXiv:1005.1776 [hep-th]].

\bibitem{Dias:2013bwa}
  \'O.~J.~C.~Dias, G.~T.~Horowitz, N.~Iqbal and J.~E.~Santos,
  ``Vortices in holographic superfluids and superconductors as conformal defects,''
  JHEP {\bf 1404}, 096 (2014)
  [arXiv:1311.3673 [hep-th]].

\bibitem{Speight}
 Johan Carlstr\"om,  Egor Babaev  and Martin Speight,
 ``Type-1.5 superconductivity in multiband systems: Effects of interband couplings,"
 Phys.\ Rev.\ B.\ {\bf 83}, 174509, (2011).


\bibitem{Tinkham}
M. Tinkham, "Introduction to Superconductivity" (Second Edition), 1996, Mac-Graw-Hill, Inc


\bibitem{annett}
James F. Annett, ``Superconductivity, Superfluids, and Condensates'', Oxford Master Series in Condensed Matter Physics, 1st Edition





\bibitem{three-band}
Y. Tanaka and T. Yanagisawa, "Chiral ground state in three-band superconductors," J. Phys. Soc. Jpn. 79, 114706 (2010). Takashi Yanagisawa, Yasumoto Tanaka, Izumi Hase, and Kunihiko Yamaji, "Vortices and Chirality in Multi-Band Superconductors", Journal of the Physical Society of Japan 81 (2012) 024712, [arXiv:1202.5400 [cond-mat.supr-con]].

\bibitem{three-band review}
For a review, see Yasumoto Tanaka, "Multicomponent superconductivity based on multiband superconductors" Supercond. Sci. Technol. 28 (2015) 034002.

\bibitem{frustrated SC}
Mitsuhiro Nishida, "Chiral ground states in a frustrated holographic superconductor", JHEP {\bf 1508} (2015) 136, arXiv:1503.00129 [hep-th].

\bibitem{hidden criticality}
L. Komendova, Yajiang Chen, A. A. Shanenko, M. V. Milo\v{s}evi\.{c}, F. M. Peeters, "Two-band superconductors: Hidden criticality deep in the superconducting state", Phys. Rev. Lett. 108, 207002 (2012), [arXiv:1203.6837 [cond-mat.supr-con]].
\end{thebibliography}
\end{document}